\documentclass[twocolumn]{aastex6}
\usepackage[]{units}
\usepackage{graphicx}
\usepackage{amssymb}
\usepackage{amsmath}
\usepackage{threeparttable}
\usepackage{hyperref}
\usepackage{color}

\newcommand{\oiii}{\hbox{[O$\,${\scriptsize III}]}}
\newcommand{\NeV}{\hbox{[Ne$\,${\scriptsize V}]}}
\newcommand{\nii}{\hbox{[N$\,${\scriptsize II}]}}
\newcommand{\mgii}{\hbox{[Mg$\,${\scriptsize II}]}}
\newcommand{\feii}{\hbox{[Fe$\,${\scriptsize II}]}}
\newcommand{\ergs}{{\rm erg\,s^{-1}}}

\mathchardef\mhyphen="2D
\slugcomment{To be submitted to MNRAS}
\shorttitle{Type 2 quasars in BOSS}
\shortauthors{Yuan et al.}

\begin{document}

\title{Spectroscopic Identification of Type 2 Quasars at $z<1$ in SDSS-III/BOSS}

\author{Sihan Yuan\altaffilmark{1}, Michael A. Strauss\altaffilmark{1}, Nadia L. Zakamska\altaffilmark{2,3}}
\altaffiltext{1}{Department of Astrophysical Sciences, Princeton University, Princeton, NJ 08544, USA}
\altaffiltext{2}{Deborah Lunder and Alan Ezekowitz Founders' Circle Member, Institute for Advanced Study, Einstein Dr., Princeton, NJ 08540, USA} 
\altaffiltext{3}{Department of Physics \& Astronomy, Johns Hopkins University, Bloomberg Center, 3400 N. Charles St., Baltimore, MD 21218, USA}

\begin{abstract}
The physics and demographics of type 2 quasars remain poorly
understood, and new samples of such objects selected in a variety of
ways can give insight into their physical properties, evolution, and
relationship to their host galaxies. We present a sample of 2758 type
2 quasars at $z\la 1$ from the SDSS-III/BOSS spectroscopic database,
selected on the basis of their emission-line properties. We probe the
luminous end of the population by requiring the rest-frame equivalent
width of \oiii\ to be $>100$\AA. We distinguish our objects from
star-forming galaxies and type 1 quasars using line widths, standard
emission line ratio diagnostic diagrams at $z<0.52$ and detection of
\NeV$\lambda$3426\AA\ at $z>0.52$. The majority of our objects have
\oiii\ luminosities in the range $1.2\times 10^{42} - 3.8\times
  10^{43}\,\ergs$ and redshifts between 0.4 and 0.65. Our sample
includes over 400 type 2 quasars with incorrectly measured redshifts
in the BOSS database; such objects often show kinematic substructure
or outflows in the \oiii\ line. The majority of the sample has
counterparts in the WISE survey, with median infrared luminosity $\nu
L_{\nu}[12\micron]=4.2\times 10^{44}\,\ergs$. Only 34 per cent of the
newly identified type 2 quasars would be selected by infrared color
cuts designed to identify obscured active nuclei, highlighting the
difficulty of identifying complete samples of type 2 quasars. We make
public the multi-Gaussian decompositions of all \oiii\ profiles for
the new sample and for 568 type 2 quasars from SDSS I/II, together
with non-parametric measures of the \oiii\ line profile shapes.  We
also identify over 600 candidate double-peaked \oiii\ profiles. 
\end{abstract}


\section{Introduction}
\label{sec:intro}

Much, if not most, of the supermassive black hole growth activity in
the Universe is hidden by gas and dust \citep{anto93, lacy15}. The
precise accounting of the demographics of active galactic nuclei
(AGNs) of different types and at different redshifts is of significant
interest because of the growing realization that the growth of
supermassive black holes may have had a strong impact on the evolution
of massive galaxies \citep{tabo93, silk98, spri05}, especially during
the obscured but intrinsically luminous (quasar) phase \citep{sand88,
  hopk06}.

Circumnuclear gas and dust make obscured (type 2) AGNs
faint at optical, ultraviolet and soft X-ray wavelengths.  But
luminous type 2 quasars ($L_{\rm bol}\ga 10^{45}\,\ergs$) may be
identified using surveys at hard X-ray \citep{norm02, bran05, hasi08,
  brus10}, infrared \citep{lacy04, ster05, mart05, lacy07, donl12,
  ster12b, eise12, glik12, lacy13, lacy15}, and radio \citep{mcca93, mart06}
wavelengths.  However, because different selection methods probe
somewhat different populations of objects, there is not yet agreement
about the obscuration fraction as
a function of redshift and luminosity 
\citep{ueda03,bran05,reye08,lawr10}, especially in pencil-beam surveys which contain very
few objects at the luminous end of the luminosity function. Very large
area surveys are important for discovering such rare 
sources, and thus despite the suppression of the apparent optical flux
by obscuration, $\sim 1000$ type 2 quasars have been selected using
their characteristic strong narrow emission lines from the
Sloan Digital Sky Survey (SDSS; \citealt{york00}), both at low ($z<1$,
\citealt{kauf03a,hao05a,zaka03,reye08,mull13}) and at high ($z\ga 2$, \citealt{alex13,
  ross15}) redshifts. 

The Baryon Oscillation Spectroscopic Survey (BOSS; \citealt{daws13})
is one of the four major surveys of the third phase of SDSS, SDSS-III
(2009-2014; \citealt{eise11}). It collected spectra of over a million
galaxies \citep{reid16} and over 300,000 quasars \citep{ross12}
selected from SDSS imaging data to measure the scale of baryon
acoustic oscillations as a function of redshift \citep{aubo15}.  The BOSS
spectrograph \citep{smee13} covers the range $3600-10400$\AA, with a
resolution of 1500-2600, depending on wavelength.  The BOSS
spectroscopic pipeline \citep{bolt12} fits the resulting spectra with
templates of common types of objects to provide redshifts and
spectroscopic classifications, and measures the
strengths and widths of various emission lines. Spectroscopic
targeting in BOSS probes fluxes $\sim 2$ mag fainter than those
accessible to the SDSS-I/II surveys \citep{daws13}, and thus one might
expect that the BOSS survey may be able to uncover a previously missed
population of optically obscured type 2 quasars.  

This paper selects type 2 quasars from the BOSS spectroscopic data. 
In Section \ref{sec:data} we describe the sample selection,
using various techniques to select $z < 0.52$ quasars (where standard
emission-line ratio selection works well) and those at higher redshift
(where the presence of the \NeV$\lambda$3426\AA\ line allows us to
distinguish AGN from star-forming galaxies).  We also identify a
significant number of type 2 quasars whose redshifts are
incorrectly measured by the BOSS pipeline. In Section \ref{sec:disc}
we discuss optical and multi-wavelength properties of the sample.  We summarize in Section
\ref{sec:conc}. We use a $h$=0.7, $\Omega_m$=0.3,
$\Omega_{\Lambda}$=0.7 cosmology throughout this paper. While SDSS
uses vacuum wavelengths, we quote emission line wavelengths in air
following established convention -- for example,
\oiii$\lambda$5007\AA\ (hereafter \oiii) has a vacuum wavelength
5008.3\AA. Objects are identified in the figures by their SDSS spectroscopic ID in the order plate - fiber - MJD. 

\section{Sample selection}
\label{sec:data}

In this paper we identify type 2 quasar candidates from the complete
SDSS-III/BOSS spectroscopic database (Data Release 12;
\citealt{alam15}). The first catalog of luminous $z\la 1$ type 2 AGNs
in the SDSS data (DR1) \citep{zaka03} was designed to be as inclusive
as possible, 
covering \oiii\ luminosities between $3.8\times 10^{40}$ and $3.8\times 10^{43}\,\ergs$,
although with completeness and selection efficiency strongly varying
with line luminosity (see also \citealt{kauf03a,hao05b}).  
In the catalog by \citet{reye08}, we set a minimal luminosity
threshold $L$\oiii$>3.8\times 10^{41}\,\ergs$, because it was not practical
to accurately measure weaker emission lines in moderate-redshift, low
signal-to-noise ratio (SNR) spectra and because we were interested in the objects
at the quasar (rather than Seyfert) end of the luminosity
range. \citet{zaka14} carried out a kinematic analysis of the
\oiii\ emission line of this sample, showing evidence for outflows
correlated with radio power and infrared luminosity.  This analysis required high SNR spectra
of the emission lines,
and thus the sample was further restricted
to luminosities $L$\oiii$>1.2\times 10^{42}\,\ergs$. 

\begin{figure}
\centering
\includegraphics[scale=0.42]{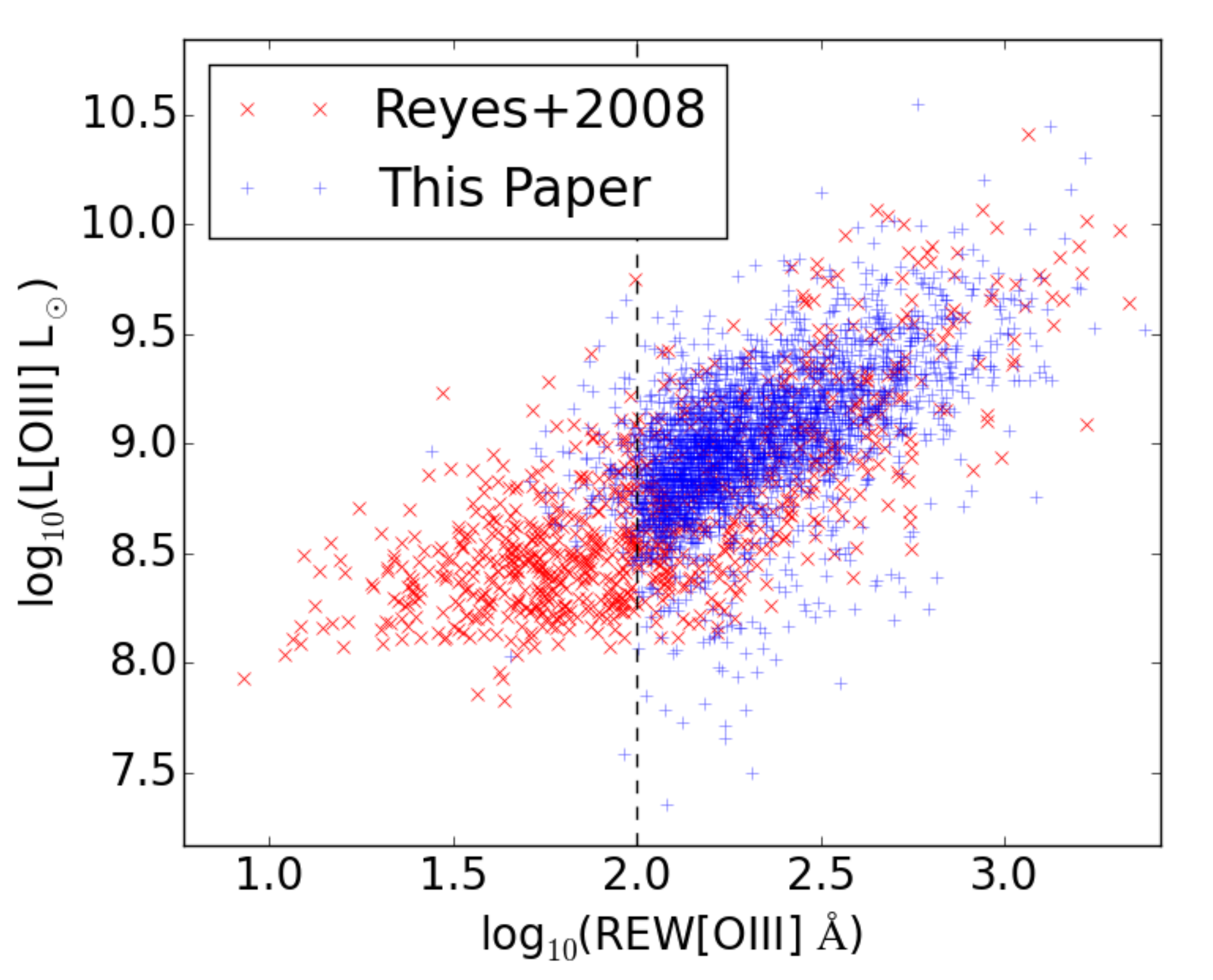}
\caption{\oiii\ rest equivalent widths and emission line luminosities
  of type 2 quasars from the sample presented in this paper (blue) and
  \citet{reye08} (red). While the sample selection is performed using
  line measurements from the BOSS pipeline, we remeasure
  \oiii\ luminosities and REWs as described in
  Section~\ref{subsec:OIII}, and these are the values shown in this
  figure. The vertical dashed line shows our selection criterion
  REW\oiii$>100$\AA\ as measured with our fits as described in
    \S~\ref{subsec:OIII}.  There is a strong positive correlation between the equivalent widths and luminosities of the \oiii\ emission line.} 
\label{fig:ew}
\end{figure}

In this paper we adopt a somewhat different approach. We rely on the
strong empirical relationship between the rest equivalent width (REW)
of the \oiii\ emission line and its luminosity (Figure \ref{fig:ew})
and aim to select type 2 quasar candidates with REW\oiii$>$100\AA\ as
completely as possible. Because \oiii\ leaves the BOSS spectral
coverage at $z \sim 1$, our sample is limited in practice to 
redshifts below unity.  The correlation in Figure~\ref{fig:ew} shows that the
\oiii\ equivalent width cut should result in a sample which is
essentially complete at $L$\oiii$>3.8\times 10^{42}\,\ergs$; 92 per cent of the objects
in \citealt{reye08} with these luminosities have REW\oiii$>$100\AA. 
Similarly 56 per cent of the
\citealt{reye08} objects with $L$\oiii\ between $1.2\times 10^{42}\,\ergs$
and $3.8\times 10^{42}\,\ergs$ have REW\oiii$>$100\AA, so we expect to include
about half of the type 2 quasars in the BOSS sample within this luminosity
range. 

The optical continuum of type 2 quasars is a poor measure of their
intrinsic power since it is suppressed by extinction, but \oiii\ 
emission is thought to arise outside of the obscuring
region, and its luminosity is correlated with bolometric luminosity
\citep{heck04}. Empirically, in type 1 quasars $L_{\rm [OIII]} =
1.2\times 10^{42}\,\ergs$ corresponds to an intrinsic (unobscured) absolute
AB magnitude of $M_{2500}=-24.0$ mag, and $3.8\times 10^{42}\,\ergs$
corresponds to $M_{2500}=-25.3$ mag \citep{reye08}. Therefore, we
estimate that the REW\oiii$>100$\AA\ criterion selects objects that
are more luminous than the traditional (though arbitrary) boundary
between Seyferts and quasars ($M_B\sim -23$ mag), at the bright end of
the quasar luminosity function at $z<1$ \citep{rich06b,hopk07}.

In this paper we define type 1 and type 2 quasars by their classical
optical signatures. Specifically, in type 1 quasars, we expect to see both
the broad-line region and the narrow-line region, whereas in type 2 quasars
the broad-line region is obscured \citep{anto93}. Therefore, the width
of the Balmer lines \citep{hao05a} and the ratio of the strengths of
the \oiii\ and H$\beta$ lines
\citep{zaka03} play a role in distinguishing type 1 from type 2
quasars. Furthermore, type 2 quasars need to be separated from
star-forming galaxies, which also have strong emission lines but with
line ratios characterized by underlying ionizing radiation produced
solely by stars. We start with objects identifiable using traditional
emission line ratio diagnostic diagrams \citep{bald81, veil87} in
Section \ref{ssec:lowz}. At redshifts $z>0.52$, the
H$\alpha+$\nii$\lambda\lambda$6548,6583\AA\AA\ emission line complex
moves out of the BOSS spectral coverage, and we develop an alternative
method that utilizes the \NeV$\lambda$ 3426\AA\ emission line to
identify type 2 quasar candidates in Section
\ref{ssec:highz}. Recovery of type 2 quasars which have been assigned
erroneous redshifts by the BOSS pipeline is described in Section
\ref{ssec:wrongz}. We present the final sample in Section \ref{ssec:sample}. 

\subsection{Selection at $z < 0.52$}
\label{ssec:lowz}

The majority of extragalactic sources with emission-line optical spectra fall into
three categories: (i) star-forming galaxies; (ii) type 1 (broad-line,
unobscured) AGNs; and (iii) type 2 (narrow-line, obscured) AGNs.
We identify 9454 emission-line objects in BOSS DR12 with 
REW\oiii$>100$\AA\ and pipeline redshift $z<1$, both as measured by
the BOSS pipeline \citep{bolt12}. For the initial selection, we use \verb+spZline+ data products\footnote{\url{http://data.sdss3.org/datamodel/files/BOSS_SPECTRO_REDUX/RUN2D/PLATE4/RUN1D/spZline.html}} for line flux measurements, but we remeasure the
\oiii\ fluxes and equivalent widths more accurately as a final step of our
catalog presentation in Section~\ref{subsec:OIII}. In this section we consider only
objects flagged 
as having confident redshift measurements by the BOSS pipeline (i.e.,
the {\tt ZWARNING} flag is set to zero; see \citealt{bolt12}), but
we return to this issue in Section~\ref{ssec:wrongz}.  
Most of the type 1 AGNs are
removed from this sample by the REW cut, as the typical REW\oiii\ in
these objects is $\sim 13$\AA\ \citep{vand01}.  

\begin{figure}
\centering
\includegraphics[scale=0.42]{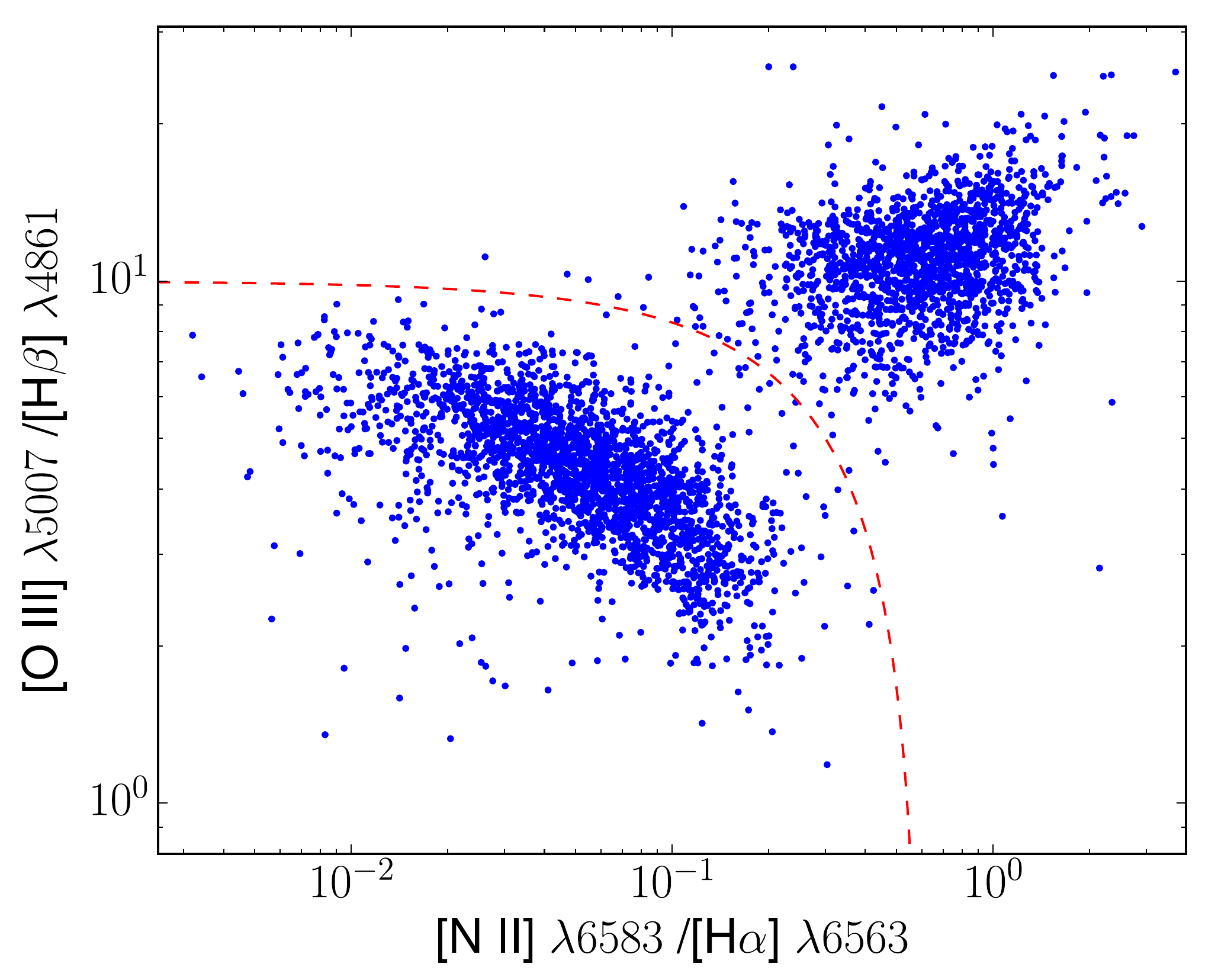}
\caption{Emission-line ratio diagnostic diagram for 4102 objects with
  $z<0.52$ and REW$>100$\AA. All measurements are from the BOSS
  pipeline.  Most type 1 AGNs are rejected by this
  equivalent width criterion, so the majority of sources in the
  diagram are type 2 AGNs and low-metallicity, low-redshift
  star-forming galaxies. The dashed line shows our type 2 AGN
  selection criterion (equation \ref{eq:bpt}).} 
\label{fig:bpt}
\end{figure}

 At $z<0.52$, for which the
 \nii$\lambda\lambda$6548,6583\AA\AA\ doublet is
   covered by the BOSS wavelength range, we use diagnostic line-ratio
   diagrams \citep{bald81, veil87, kewl01, hao05a} to separate type 2
   AGNs from star-forming galaxies. Our sample of 9454 high REW
   objects includes 4102 objects with redshift $z<0.52$ and with
   $>3\,\sigma$ detections of the relevant lines as measured by the
   BOSS pipeline: \nii$\lambda$6583\AA,
   H$\alpha\lambda$6563\AA, \oiii$\lambda$5007\AA\ and
   H$\beta\lambda$4861\AA. The distribution of
   these objects in the \oiii/H$\beta$ vs \nii/H$\alpha$
   plane is shown in Figure \ref{fig:bpt}. At these high equivalent
   widths, most of the star-forming galaxies tend to have relatively
   low metallicities because \oiii\ is one of the few available
   coolants of the low-metallicity gas. These galaxies separate
   cleanly from the AGNs \citep{kewl01,kauf03a,hao05a}; at
   this equivalent width and luminosity, objects lying between the
   star-forming and AGN branches would have to have unusually high
   luminosity in both components, which is quite rare.
We select AGNs with the cut shown in Figure \ref{fig:bpt}:
 
\begin{equation}
\frac{\rm \oiii\lambda 5007\AA}{\rm H\beta}>10.0-16.7\times \frac{\rm \nii\lambda 6583\AA}{\rm H\alpha}.
\label{eq:bpt}
\end{equation}
This cut further reduces the type 1 AGN contamination of the
sample, as \oiii/H$\beta$ tends to be low in type 1 objects, where
H$\beta$ is dominated by the broad component. We find 1693 type 2
quasar candidates at $z<0.52$ which have REW\oiii$>100$\AA\ and
which are above the diagnostic cut given by equation (\ref{eq:bpt}). A
visual inspection of these candidate yields a final list of 1606 type
2s at $z<0.52$, after removing a modest number of objects with weak
broad-line components that the BOSS pipeline failed to model.  

\subsection{Selection at $z>0.52$} 
\label{ssec:highz}

Various methods have been suggested for separating the spectra of type 2 AGNs from star-forming galaxies at redshifts where the full set of diagnostic emission lines is not available \citep{zaka03, reye08, gill10}. In our case, we need to develop new selection criteria to identify type 2 quasar candidates at $z>0.52$, when \nii$\lambda$6583\AA\ moves out of the BOSS wavelength range. 

We use the \NeV$\lambda$3426\AA\ emission line to distinguish AGN from
star-forming galaxies \citep{zaka03, gill10}.  The ionization energy
of Ne$^{4+}$ is 97 eV, so the presence of this line implies that the
gas has been ionized with intense radiation in the hard UV and soft
X-ray range. Star formation produces essentially no emission at these
wavelengths, so the mere detection of the \NeV\ emission is an
unambiguous sign of the presence of an AGN.  

\begin{figure}
\centering
\includegraphics[scale=0.35]{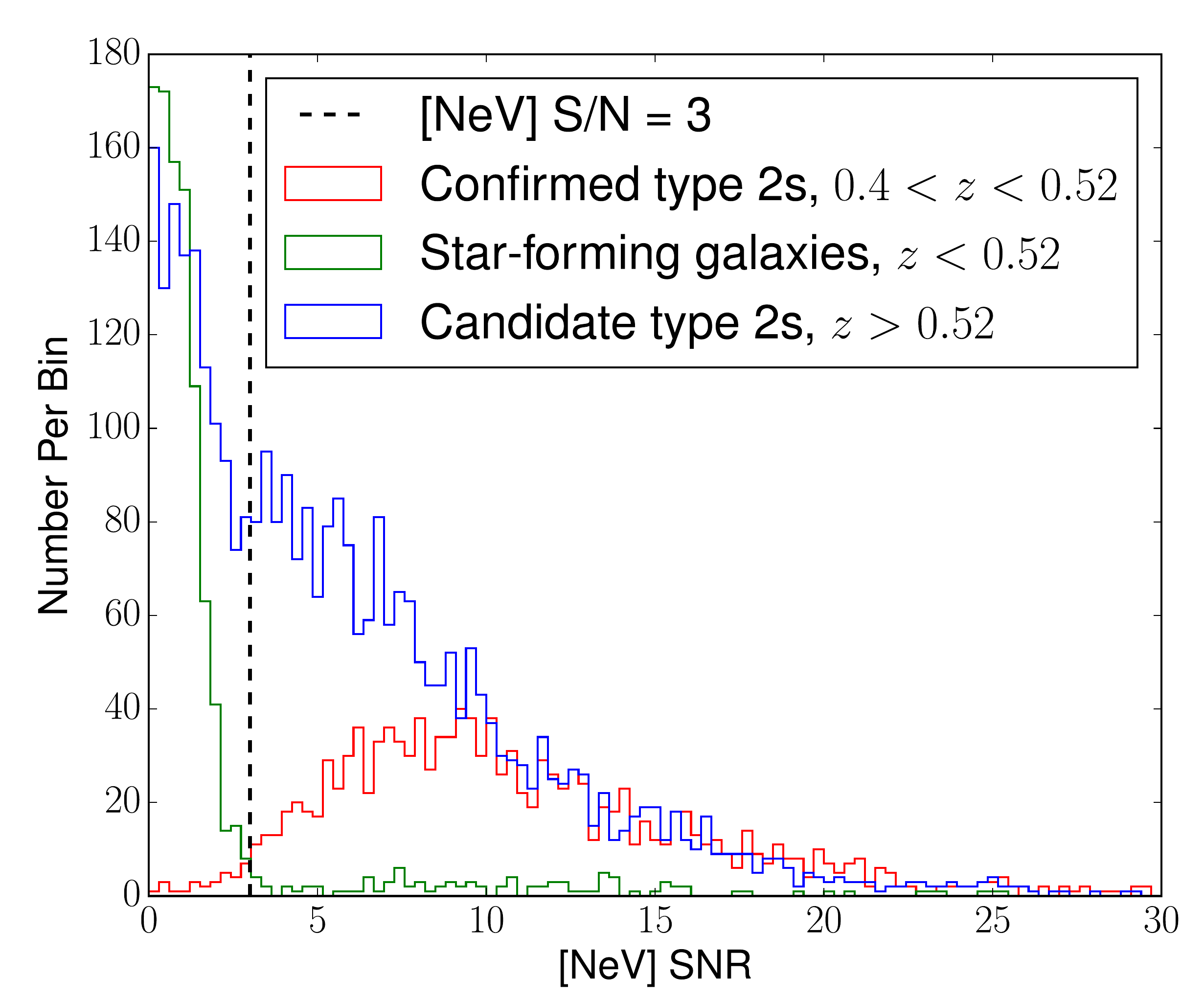} 
\caption{\NeV\ SNR distribution of type 2 quasars at $0.4<z<0.52$ (red
  histogram), type 2 quasar candidates at $z > 0.52$ (blue histogram)
  and a subset of the star-forming galaxies at $z<0.52$ (green
  histogram). The dashed line marks the SNR${}> 3$ cutoff that we
  choose for our quasar selection for objects with $z > 0.52$. While
  approximately 98 per cent of confirmed type 2 quasars at $0.4< z <0.52$
  show \NeV\ SNR${}> 3$, over 91 per cent of star forming galaxies show
  no \NeV\ detection with SNR${}> 3$.}
\label{fig:neon}
\end{figure}

The BOSS pipeline does
not automatically measure \NeV\ fluxes, so we measure them using
single-Gaussian fits in all candidate spectra. To test how well such
\NeV-based selection performs, we measure \NeV\ fluxes in AGN and
star-forming galaxies with $0.40 < z < 0.52$ from the sample shown in
Figure~\ref{fig:bpt}. The classification of these objects is already
known from the standard diagnostic diagrams. The distribution of the
SNR of the \NeV\ line is shown in Figure~\ref{fig:neon}: only 9 per cent of
the objects classified as star-forming by the line ratio criterion of
equation~(\ref{eq:bpt}) have \NeV\ detections; the vast majority
  of the objects classified as star-forming with \NeV\ detections turn
  out to be Type 1 objects, 
  with broad bases to the H$\alpha$ lines.  The few exceptions all lie
  close to the boundary between the star-forming and AGN branches in
  the emission line ratio diagram (Figure~\ref{fig:bpt}), and thus are
  likely an admixture of the two components.

  On the other hand, 98 per cent of
  the BPT-selected AGNs show significant \NeV\ emission. Thus the mere
detection of the \NeV\ line is indeed a good indicator of the presence
of an AGN and results in a fairly complete sample.  

There are 4143 objects in BOSS DR12 with $z>0.52$ and
REW\oiii$>100$\AA. For a source to be selected as a type 2 quasar
candidate, we require that the \NeV\ line be detected at SNR${}>3$ in
the BOSS spectra for $z>0.52$ objects with REW\oiii${}>100$\AA.  2191
objects satisfy these criteria. The SNR criterion is of course
dependent on the SNR of the BOSS spectra themselves, and we find that
the median spectral SNR per pixel of the high equivalent width objects
drops 
steadily from 4 at $z=0.4$ to 2.5 at $z>0.6$. Thus 
our \NeV-based selection likely becomes less complete at high
redshifts. 

\begin{figure}
\centering
\includegraphics[scale=0.42]{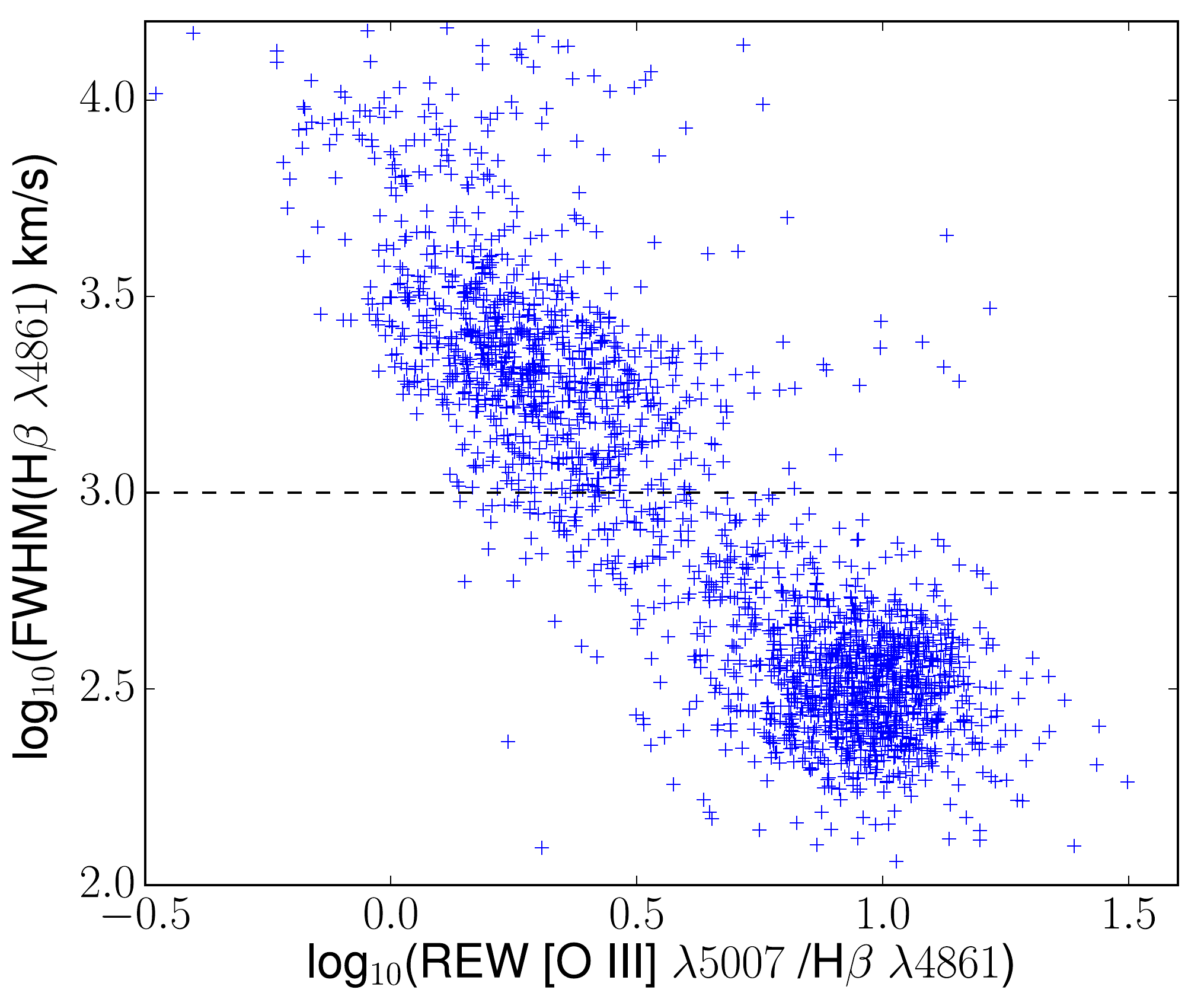} 
\caption{The relationship between FWHM(H$\beta$) and \oiii/H$\beta$
  for 2191 AGNs with $z > 0.52$,
  REW\oiii$>$100\AA, and $>3\,\sigma$ detection of the
  \NeV\ line. All measurements are from the BOSS pipeline.  The objects in
  the bottom right quadrant have narrow H$\beta$ lines and high
  \oiii/H$\beta$ ratios and are likely type 2 AGNs. The 
  objects in the top left have broad H$\beta$ lines and low
  \oiii/H$\beta$ ratios (presumably dominated by the broad H$\beta$
  component) and are mostly type 1 AGNs. We choose FWHM(H$\beta$)$ =
  $1000 km s$^{-1}$ (dashed line) as our selection cut to remove
  broad-line type 1 AGNs.  Visual inspection found only 10
  objects with broader H$\beta$ lines which belong in the type 2
  category.}
\label{fig:hbfwhm}
\end{figure}

The sample at this stage still includes a substantial number of
type 1 AGN. Figure~\ref{fig:hbfwhm} shows the full width at half maximum
(FWHM) of the H$\beta$ line against the ratio of \oiii\ to H$\beta$
for the 2191 objects. The sample is clearly bimodal in H$\beta$ line
width, which is used as a classical distinguishing characteristic
between type 1 and type 2 AGNs at low luminosities \citep{khac74,
  hao05a}. We choose FWHM(H$\beta) = 1000$ km s$^{-1}$ as the cut-off
to remove broad-line type 1 AGNs. With this cut, we select 1250 type 2
candidates with $z > 0.52$, REW\oiii\ $>100$\AA, \NeV\ SNR $> 3$ and
FWHM(H$\beta$) $<$1000 km s$^{-1}$. A visual inspection yields 796 type
2 quasars. Most of the candidates rejected by this visual inspection
showed weak broad H$\beta$, broad \mgii$\lambda 2800$\AA\ or a strong
blue continuum (the latter often associated with narrow-line Seyfert 1
galaxies; see \citealt{will02}).  

In the presence of strong quasar-driven outflows, the kinematics of
the forbidden-line region can sometimes result in FWHM of the extended
emission line region in excess of 1000 km s$^{-1}$.  Indeed,
  blueshifted asymmetries of the \oiii\ line have been recognized as
  the signature of outflows since the 1980's
  \citep{heck81,dero84,whit85a,wils85}, although the relationship
  between outflows and FWHM significantly larger than the depth of 
  the galaxy potential well became clear only much more recently
  \citep{nesv06,nesv08,gree09,gree11,vill11a,hain13,zaka14,zaka16b}.
 To avoid missing the sources with the highest FWHM, we visually inspect the 941 objects in Figure
\ref{fig:hbfwhm} above the FWHM(H$\beta$)$=$1000 km s$^{-1}$ cutoff
line. We identify an additional 10 type 2 quasars. This is only a
small fraction of the strongly kinematically disturbed type 2 quasars
in our sample, most of which are identified in the BOSS catalog
using another method (Section \ref{ssec:wrongz}). Our final sample
from \NeV-based selection at $z>0.52$ includes 806 type 2 quasars with
$z > 0.52$.  

It is possible that we have been overly aggressive in rejecting
objects with weak broad components in H$\beta$ or
\mgii$\lambda$2800\AA. The problem of weeding out type 1 AGNs with
weak broad lines from genuine type 2 candidates is inherently
difficult. Even when the direct lines of sight to
the nucleus are obscured, some quasar emission can escape along other
directions, scatter off the interstellar medium of the host galaxy and
reach the observer \citep{anto85,anto93,zaka05}. If the scattering is
more efficient than a few percent, then this component can make a
noticeable enough contribution to the integrated spectrum of the
object that we would see weak broad components in the Balmer and
\mgii\ lines as well as a continuum rising to the blue. Short of
conducting polarimetry or spectropolarimetry, we cannot distinguish
such objects from type 1 AGN with weak lines and weak continuum. Thus
our selection procedure in which we reject all objects with detectable
broad components unfortunately biases our sample against 
type 2 quasars with high scattering efficiency. 

Another problem is that some genuine type 2 quasars show narrow features near \mgii. In \citet{zaka05}, He {\scriptsize II} $\lambda$ 2734\AA\ and C {\scriptsize II} $\lambda$2838\AA\ (vacuum wavelengths) are tentatively identified as possible satellite features to \mgii. In lower SNR data or in an object with high velocities in forbidden lines, these features could be blended together and be erroneously interpreted as a broad Mg component. We keep objects without other indications of being type 1 quasars, if the satellite lines are clearly spectroscopically resolved from \mgii.

Finally, in unobscured AGNs the region near Mg has strong emission from multiple lines of \feii.  In particular, narrow-line Seyfert 1 galaxies show a broad Fe complex peaking at 2300-2400\AA\ and two Fe complexes on either side of \mgii\ \citep{cons03}. Thus Mg can appear as a narrow core with broad ``shoulders'' which are actually Fe complexes. Even if such objects show no other signatures of being unobscured, we reject them from our sample. Because we require a high REW of \oiii, a narrow H$\beta$, and low \feii\ during the visual inspection stage we expect little to no contamination of our sample by narrow-line Seyfert 1 galaxies. 

\subsection{Selection of type 2s with incorrect/unreliable redshifts}
\label{ssec:wrongz}

Even though 99.8 per cent of the redshifts the BOSS pipeline flags as
reliable are correct \citep{adel08, bolt12},
there are still some objects with incorrect redshifts in the
database. Type 2 quasars can be among 
those mis-classified objects because there is no proper template for
them in the BOSS pipeline. To identify such objects, we explore three
ways in which the pipeline is known to respond erroneously to a
type 2 spectrum. First, the strong \oiii\ emission line of type 2 quasars
could be mis-identified as Ly$\alpha$, resulting in a mistakenly high
redshift. For example, mis-identification of \oiii$\lambda$5007\AA\ at
redshift $z_{\rm true}=0.5$ as Ly$\alpha$ would yield $z_{\rm
  wrong}=5.18$. Second, the \oiii\ emission line could be mis-identified 
as H$\alpha$, which would result in a mistakenly low redshift. For
example, mis-identification of \oiii$\lambda$5007\AA\ at $z_{\rm
  true}=0.5$ as H$\alpha$ would yield $z_{\rm wrong}=0.114$.  Finally,
the redshift could be measured correctly, but the pipeline could
indicate that it has low confidence in the result.  

\begin{figure}[h]
\centering
\includegraphics[scale=0.45]{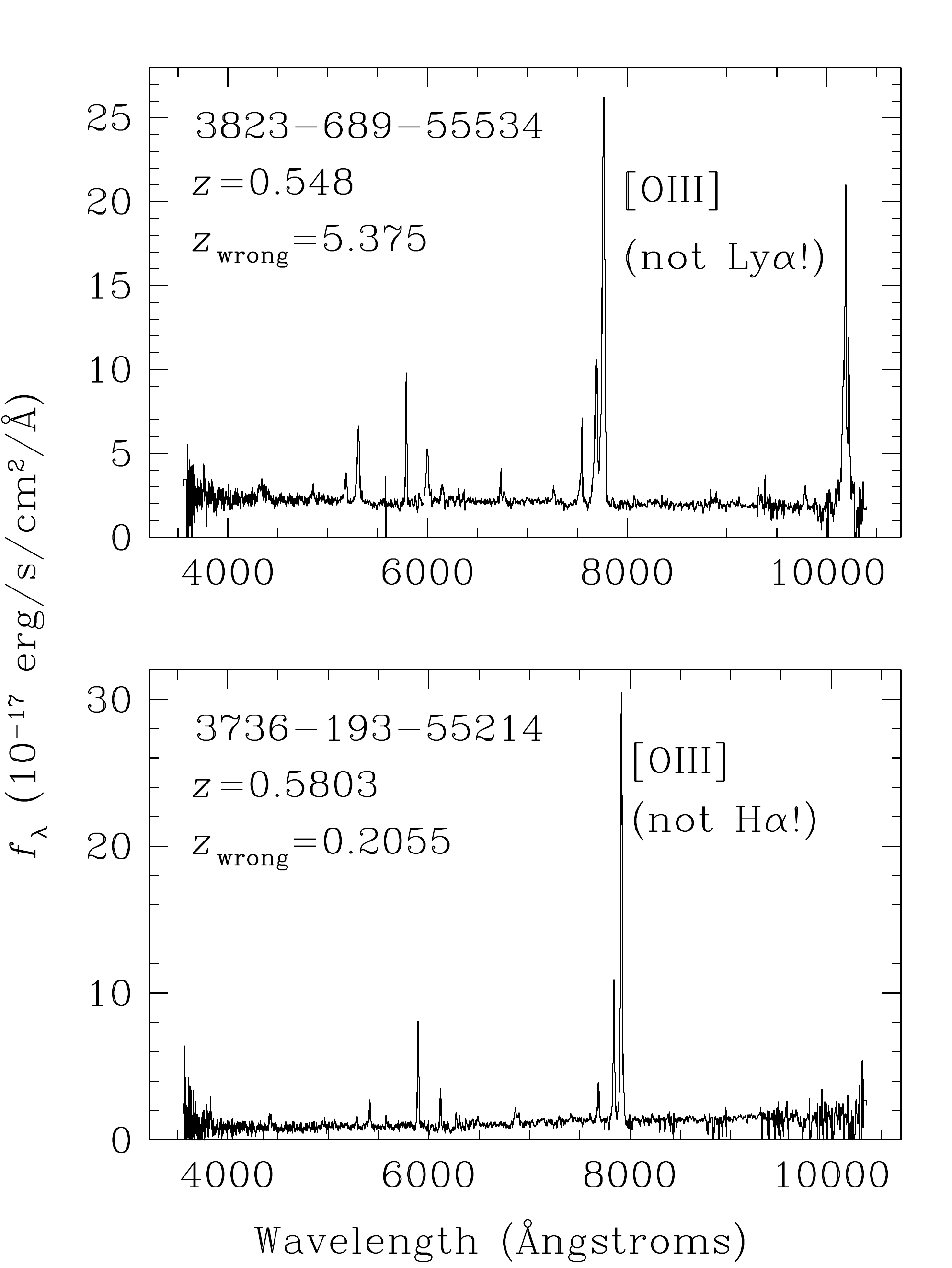}
\caption{Example type 2 quasars identified assuming that the BOSS
  pipeline mistook \oiii\ for another strong emission line: for
  Ly$\alpha$ in the top panel (true redshift $z_{\rm true}=0.548$) and
  for H$\alpha$ in the bottom panel (true redshift $z_{\rm
    true}=0.5803$). Each quasar is indicated with its plate, fiber, and
  Modified Julian Date (MJD (see \S~\ref{ssec:sample}).  These
  spectra have been smoothed with a five-pixel boxcar.}  
\label{fig:wrong}
\end{figure}

In the first possibility, an \oiii\ line at $z=0$ (i.e., at 5007\AA)
misinterpreted as Ly$\alpha$ will be assigned a redshift 3.12.  We
matched the list of 26,489 BOSS objects with $z>3.12$ (as measured by the
BOSS pipeline) against the visually inspected (type 1) BOSS quasar
catalog \citep{paris15}; objects that match are presumed to have
correct redshifts.
Visually inspecting the 1715 objects which remain yields 61
type 2 quasar candidates, whereas the rest are mostly genuine high
redshift type 1 AGNs. All 61 type 2 candidates show strong
\oiii\ emission. An example of a type 2 quasar selected using this
method is shown in the top panel of Figure \ref{fig:wrong}. 

In the second possibility, \oiii\ is mis-identified as H$\alpha$,
which is only possible for $z_{\rm true}>0.31$. The conversion from
observed to rest-frame equivalent width implies that our desired cut of
REW\oiii$=$100\AA\ corresponds to a listed REW for the line,
interpreted wrongly as H$\alpha$, of 131\AA.  An object with H$\alpha$
with such a high equivalent width is likely to also exhibit strong
\oiii.  We thus looked for objects with listed H$\alpha$ REWs greater
than 131\AA, but with REW\oiii\ less than 5\AA.  This yields 625
objects, of which 369 (well over half) are in fact type 2 quasars at
the wrong redshift. Most of the remaining candidates are 
artifacts with noisy spectra. An example of a type 2 quasar from this
selection method is shown in the bottom panel of Figure \ref{fig:wrong}.  

For all selection methods presented so far, we require that the
BOSS pipeline be confident about the redshift: the \verb+ZWARNING+
flag \citep{bolt12} must be
set to 0. We select another interesting subsample of 1050 type 2s,
identifying those objects that the BOSS pipeline flags as having
problematic redshift measurements, i.e., non-sky fibers with
\verb+ZWARNING!=0+ with measured REW\oiii${}>100$\AA. 
We visually inspect this subsample and identify
78 type 2s. The BOSS pipeline redshift of these type 2s is essentially
always right, despite the \verb+ZWARNING+ flag (indeed, had the
redshifts been wrong, the REW\oiii\ measurement would have been
meaningless). 

The 508 type 2 quasar candidates selected in this section are
particularly interesting because they tend to show strongly disturbed
kinematics in their \oiii\ emission lines, which is presumably
why their redshifts are either mis-identified by the pipeline or the
pipeline is not confident about the redshift. This also explains why
we see such a clean separation of FWHM(H$\beta$) in Section
\ref{ssec:highz} and Figure~\ref{fig:hbfwhm}: the majority of strongly
kinematically disturbed objects with broad forbidden lines are placed
at a wrong redshift and are therefore not correctly identified using
the \NeV\ SNR cut employed in that section. 

\subsection{BOSS catalog of type 2 quasars}
\label{ssec:sample}

We now have 1606 spectroscopic observations of type 2 quasars from Section
\ref{ssec:lowz}, 806 from Section \ref{ssec:highz} and 508 from
Section \ref{ssec:wrongz}, adding up to a total of 2920 unique
spectroscopic observations. Accounting for objects with multiple
spectroscopic observations, our sample represents 2758 unique sources. We provide the full catalog as an online FITS table (\url{http://zakamska.johnshopkins.edu/data.htm}). The complete data structure of the catalog is described in Table \ref{tab:cat}.  

Our basic identification method is by the BOSS Plate and Fiber Number on which this object was observed spectroscopically, together with the Modified Julian Date (MJD) of the spectroscopic observation. We also provide right ascension and declination, as measured by the SDSS \citep{DR8,DR9}. In the cases when the same object has multiple spectra in the database, we flag the first appearance of the source with a `unique' flag and any subsequent appearances are flagged with the spectroscopic identification of the first available spectrum. The catalog includes \NeV\ emission line measurements described in Section \ref{ssec:highz}, \oiii\ emission line measurements described in Section \ref{subsec:OIII} and infrared luminosity measurements described in Section \ref{subsec:allwise}. 

\section{Properties of the BOSS type 2 quasars}
\label{sec:disc}

\subsection{Optical colors and target selection}

\begin{figure}
\centering
\includegraphics[scale=0.42]{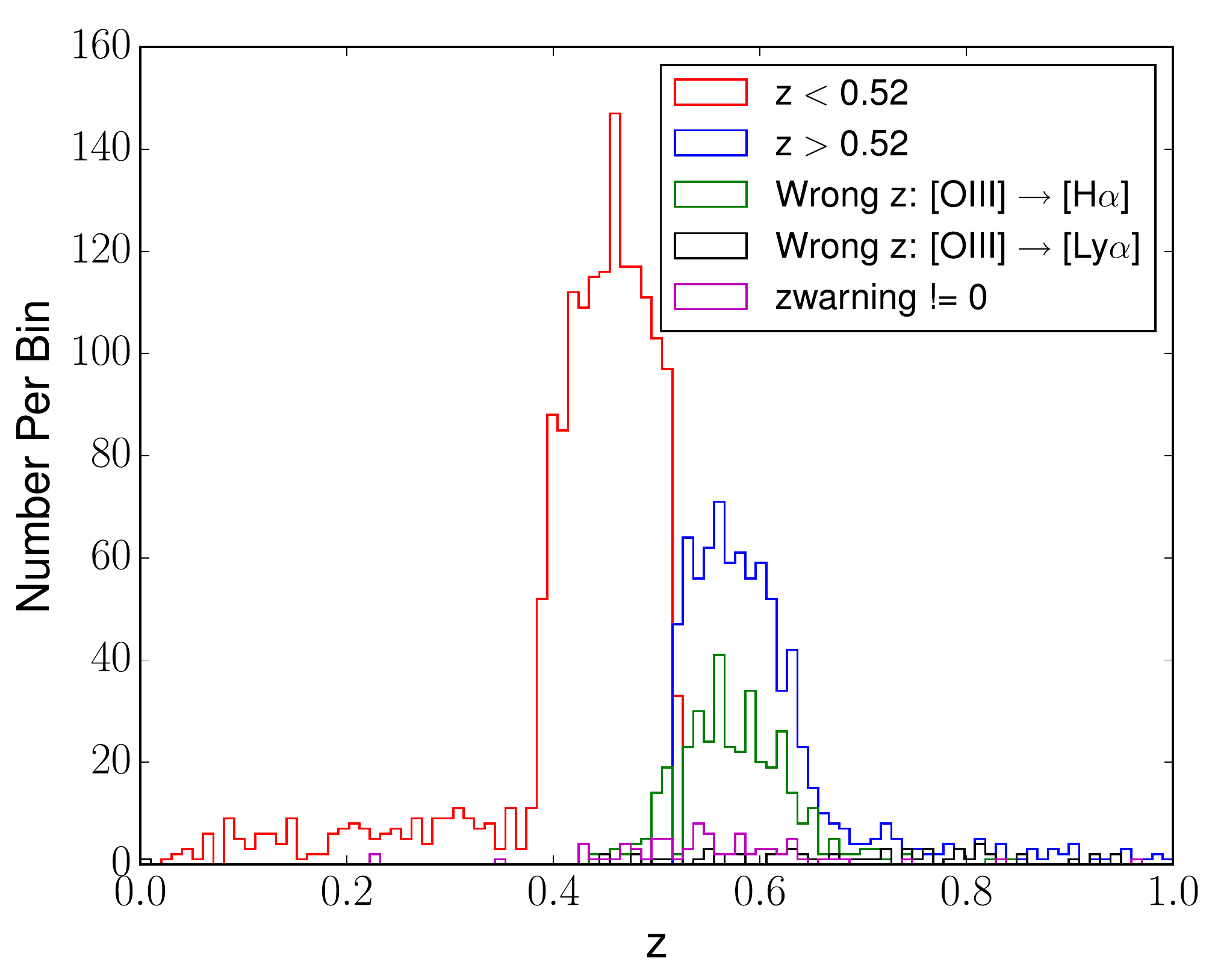}
\caption{The redshift distribution of our type 2 quasars in $\Delta z=0.01$ bins, showing the
  five selection criteria. The red histogram corresponds to the 1606
  type 2 quasars from Section \ref{ssec:lowz}; the blue histogram is
  the 806 type 2 quasars from Section \ref{ssec:highz}, and the green,
  black and magenta histograms correspond to the two samples of
  wrong-redshift-selected type 2 quasars and the {\tt ZWARNING}
  selected sample in Section \ref{ssec:wrongz}. All redshifts are as
  measured by the pipeline described in Section~\ref{subsec:OIII}.
  \label{fig:zdistri}}
\end{figure}

We present the redshift distribution of each of our subsamples in
Figure \ref{fig:zdistri}. The majority ($\sim 85$ per cent) of the
objects in the final sample lie within the redshift range $0.4 < z <
0.7$.  This is a reflection of how these objects were selected for
spectroscopy in the BOSS survey; 2480 of the type 2 quasars in the
final catalog were targeted as CMASS galaxies. These are selected with a series of magnitude and
  color cuts, as described by
\citet{reid16}, to be a roughly stellar
  mass-limited sample of galaxies (CMASS stands for ``Constant
  Mass'').

\begin{figure}[t]
\includegraphics[width=9cm]{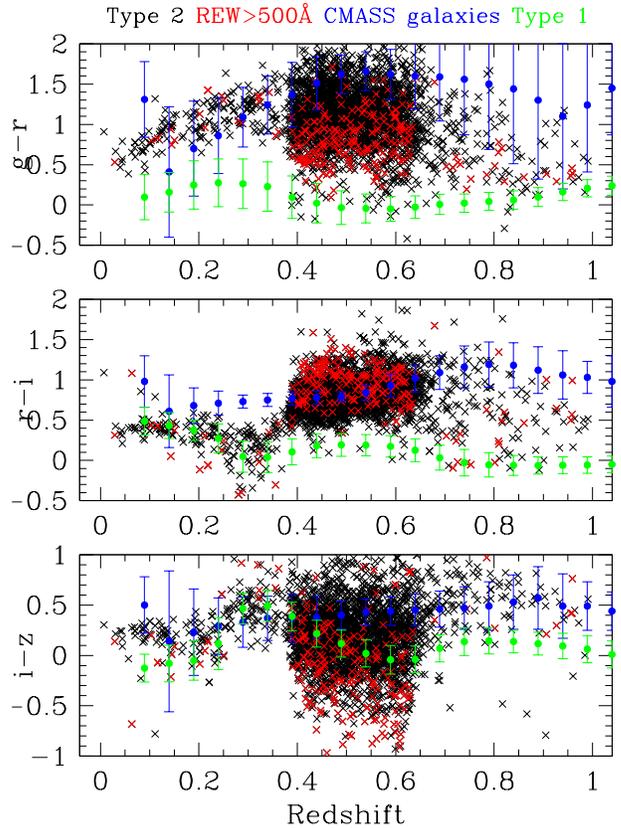}
\caption{The observed SDSS colors of objects in our sample as a
  function of redshift.  The red points correspond to those objects
  with the highest \oiii\ equivalent widths.  The blue points give the
  median colors of galaxies selected by the CMASS algorithm
  \citep{reid16}, while the green points are the median for quasars
  from SDSS-I/II \citep{schn10}. 
\label{fig:color_redshift}}
\end{figure}

Figure~\ref{fig:color_redshift} shows the SDSS measured colors
(measured using model magnitudes corrected for Galactic extinction
following \citealt{schl98}) of the 
objects in the sample as a function of redshift, following
\citet{rich03}.  Also shown are median colors for CMASS galaxies in
general, as well as for type 1 quasars from SDSS-I/II \citep{schn10}.  The
type 2 quasars in our sample tend to be appreciably bluer than the bulk of CMASS
galaxies in $g-r$ and $i-z$, but relatively red in $r-i$. Some of this interesting behaviour is due to the high equivalent widths of the emission lines in type 2 quasar spectra. Specifically, the \oiii\ line falls in
the $i$ band from redshift 0.4 to 0.6, making the $r-i$ colors relatively red, close to those of CMASS galaxies, and $i-z$ colors blue, close to or even bluer than those of type 1 quasars. 

The continuum color of our sample is bluer than that of CMASS
galaxies, as seen in $g-r$ colors which are not dominated by emission
lines, though they are not as blue as those of type 1 quasars. The dominant
contribution to the rest-frame ultraviolet continuum of type 2 quasars
is likely to be scattered light \citep{zaka05, zaka06, obie16}, though
star formation is also possible \citep{zaka16a, wyle16}. Type 2 quasar
hosts are known \citep{liu09} to be even more strongly star-forming than type 1
quasar host galaxies which appear with young stellar populations at
these redshifts \citep{kirhakos99,matsuoka14,matsuoka15}.   

 The fact that $\sim85$ per cent of our type 2 quasars are selected by the BOSS
 galaxy target selection algorithms means that most of our type 2
 quasars are resolved in SDSS imaging. This result is consistent with
 \citet{reye08};  $\sim 50$ per cent of their type 2 quasar sample were
 selected by the main galaxy target selection algorithm
 \citep{strauss02}. It also suggests that there may be an additional
 population of unresolved type 2 quasars yet to be identified which
 reside in compact galaxies; such objects would not be selected by the
 CMASS targeting algorithm and therefore would not have BOSS spectra. 

\subsection{Refitting the \oiii\ profile}
\label{subsec:OIII}

Our sample relies strongly on \oiii\ measurements and is limited in REW\oiii. We initially use the measurements
of this quantity from the BOSS pipeline outputs, but these are not
ideal: the pipeline fits a single Gaussian to the line, and forces the
width of that Gaussian to be the same for all forbidden lines fit.
Following \citet{zaka14}, we refit the \oiii\ doublet over the rest
wavelength range 4910\AA--5058\AA\ for all our objects, assuming a
2.996:1 intensity ratio for the two \oiii\ lines, and forcing the
redshifts and profiles of \oiii$\lambda$4959\AA\ and
\oiii$\lambda$5007\AA\ to be the same. Unlike the extreme objects
discussed in \citet{zaka16b}, in no case is the \oiii\ line broad
enough to be affected by the H$\beta$ line, and we thus do not include
it in the modeling.  The SNR of the spectra in these strong lines is
adequate to allow detailed fits.  Fits are carried out assuming a
linear continuum, and one, two, three or four Gaussians; if adding an
extra Gaussian component leads to a decrease in reduced $\chi^2$ of
$<10$ per cent, we accept the fit with a smaller number of components.  The
vast majority of the fits require two or three Gaussians; there are
only two sources that require four.
  
  These fits give accurate measurements of \oiii\ luminosities and
  equivalent widths; we used these results in Figure~\ref{fig:ew}.
  Because these fits are sensitive to wings in the profile that are
  not fit by a single Gaussian, they tend to give equivalent widths
  which are systematically higher than those measured by the BOSS
  pipeline. Specifically, the median REW\oiii\ for our type 2 sample as
  measured by the pipeline is 165\AA\ whereas our refits give a median
  REW\oiii\ of 186.0\AA. This means that our sample is somewhat
  incomplete close to the REW\oiii\ limit of 100\AA; there is
  presumably a population of objects with pipeline equivalent widths
  somewhat below 100\AA, which would move above this limit with the
  detailed fits we have described here. 

We provide the complete multi-Gaussian decomposition for all sources
in the catalog, as well as the 568 objects
with $L>1.2\times 10^{42}\,\ergs$ in the \citet{reye08} catalog
\citep{zaka14}, as online FITS tables (\url{http://zakamska.johnshopkins.edu/data.htm}). 
The names of the columns in the kinematic catalog are listed
in Table~\ref{tab:kin}. In addition to tabulating the Gaussian
components, we provide non-parametric measures of the \oiii\ profiles,
including the full widths at 25 per cent and 50 per cent of the maximum (in km
s$^{-1}$) and the non-parametric measures defined by \citet{zaka14}
following \citet{whit85a}. Specifically, for every emission line
profile we measure the velocities $v_x$ at which $x$ per cent of line power
accumulates. This allows us to define the widths encompassing 50 per cent,
80 per cent and 90 per cent of line power: $w_{50}\equiv v_{75}-v_{25}$,
$w_{80} \equiv v_{90}-v_{10}$ and $w_{90} \equiv v_{95}-v_{05}$, measured in km
s$^{-1}$. These widths are more sensitive to weak broad components
than the traditional full width at half maximum measures. 
We also tabulate the dimensionless relative asymmetry
$R\equiv ((v_{95}-v_{50})-(v_{50}-v_{05}))/w_{90}$, which is negative for
profiles with a heavier blue-shifted wing, and dimensionless kurtosis
$r_{9050}\equiv w_{90}/w_{50}$ which is larger for profiles with heavy wings
and narrow cores.  

\begin{figure}[t]
\includegraphics[width=9cm]{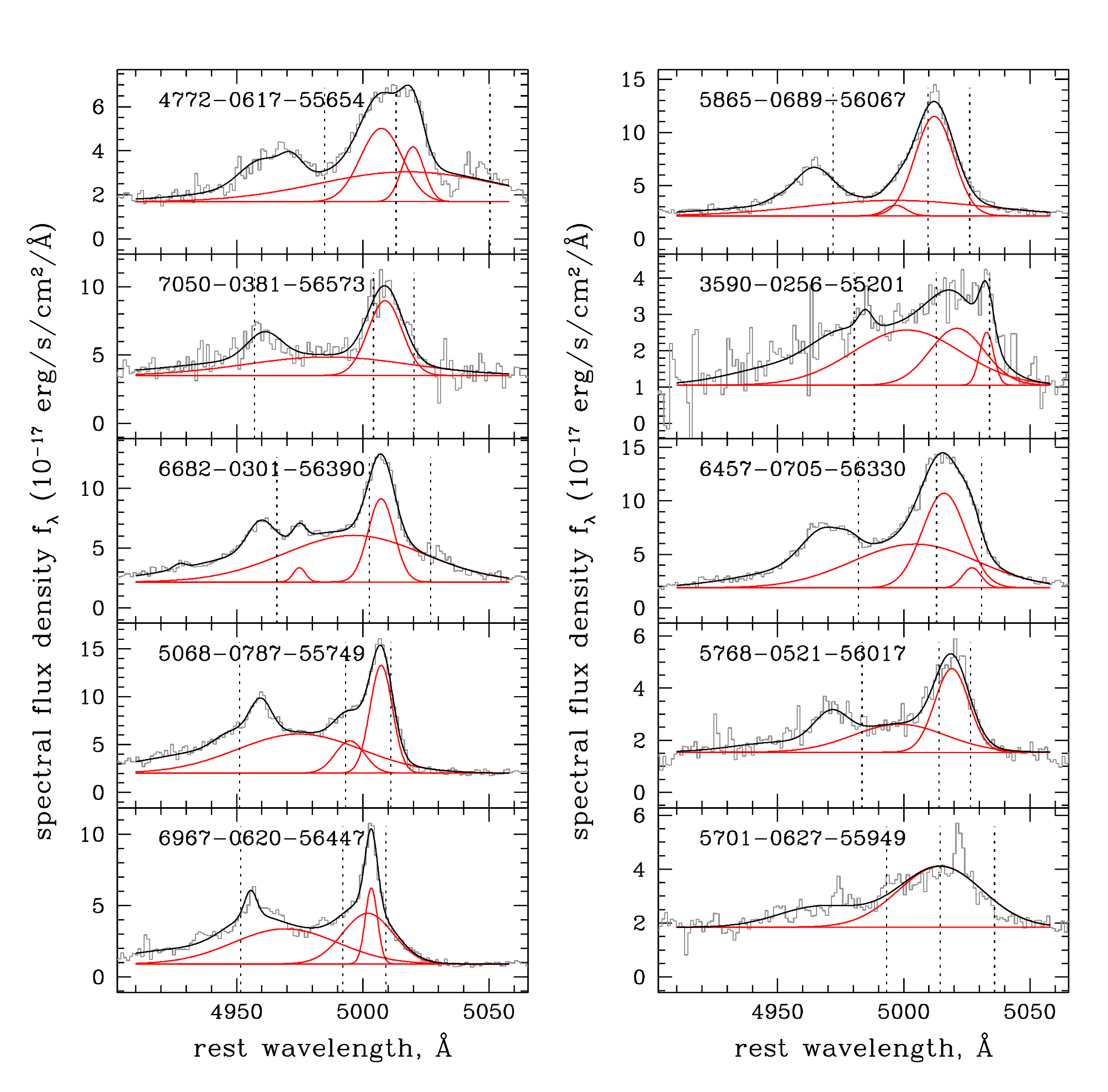}
\caption{\oiii\ spectra of the ten objects with the broadest
  \oiii\ emission, as measured by the width containing 80 per cent of the
  line power $w_{80}$. The spectrum of the
  \oiii$\lambda$4959,5007\AA\ region is shown in grey, and the model
  components (continuum and two or three Gaussians) are shown in red for the
  5007\AA\ line (the fit is performed simultaneously on both lines in
  the doublet assuming the same kinematic structure).  The summed
  model for both lines of the  \oiii\ doublet is shown in
  black. Vertical dashed lines show $v_{10}$, $v_{50}$ and $v_{90}$,
  so that $w_{80}$ is encompassed between the left and right
  lines. The ten objects shown in the figure have $w_{80}$ values in
  the range 2558 -- 3912 km s$^{-1}$. These values are comparable to
  the maximal widths found in the \citet{reye08} sample
  \citep{zaka14}, but fall short of the extreme values (up to 5409 km
  s$^{-1}$) found in high-luminosity red quasars at high redshifts
  \citep{zaka16b}. Each spectrum is labeled with its plate, fiber, and
  MJD.  
\label{fig:linefits}}\end{figure}

A comprehensive study of the relationships between these measures and their relevance to quasar winds is presented by \citet{zaka14}. In what follows we use $w_{80}$ as a measure of the \oiii\ kinematics. Figure~\ref{fig:linefits} shows the profiles and fits for those objects with the objects with the most dramatic outflows, as measured by $w_{80}$. 

In both catalogs, we flag candidate double-peaked
\oiii\ profiles (see \citealt{liu10a}). While most of these profiles likely result from
biconical quasar-driven winds, where each plowed shell may appear as a
separate Gaussian component \citep{gree12, harr15}, a small fraction
of these objects could be due to kpc-scale binary active nuclei
\citep{liu10b, shen11b, come12}. Separating these two possibilities is not yet
possible without extensive follow-up observations, so we identify
possible double-peaked emitters in the catalog exclusively based on
the shape of the \oiii\ line. While there is no formal definition of
what constitutes a double-peaked profile, for identifying candidate
kpc-scale binaries we are interested in profiles with two distinct
narrow components in the \oiii\ profile which are kinematically
separated from one another by an amount comparable to their velocity
dispersion. 

\begin{figure}
\includegraphics[width=9cm]{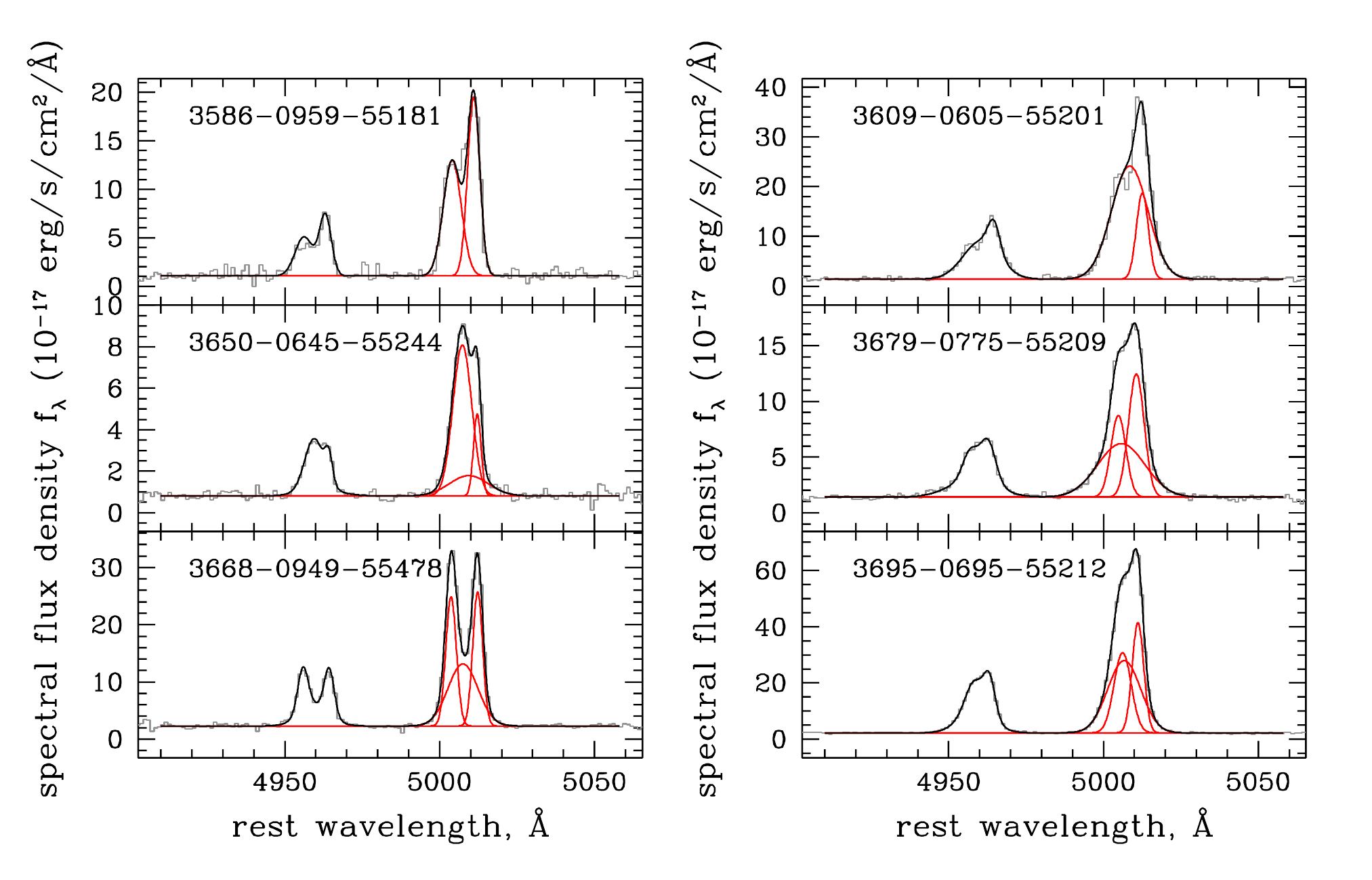}
\caption{Six of the double-peaked candidates in the catalog. The
  three objects in the left panels are selected by requiring that the
  fitted profile has a minimum, whereas the three objects in the right
  panels are selected by visual inspection.  Each spectrum is labeled
  by its plate, fiber, and MJD.} 
\label{fig:double}
\end{figure}

We identify candidate double-peaked profiles in two
ways. First, following \citet{liu10a} we identify sources with a
minimum in the fitted \oiii\ profile and examine them visually. Most of these objects are retained as double-peaked candidates. Second,
we visually examine all profiles and flag those that do not have a
minimum but nonetheless appear to have two distinct kinematic
components. Candidates are visually flagged based on the observed profiles, regardless of whether the two distinct components are accurately captured by the multi-Gaussian fits. Examples of objects from both selection methods are shown
in Figure \ref{fig:double}. In the new BOSS sample, 420 of the 2920
spectra show a profile minimum. Of these, 363 (86 per cent) are retained as double-peaked after visual inspection, and an additional 183 without a model profile minimum are identified visually. In the \citet{zaka14} sample, there are, respectively, 60 and 48 double-peaked candidates identified by the two algorithms. 

\subsection{Crossmatch with AllWISE}
\label{subsec:allwise}

The dust within the obscuring medium along the line of sight to a type 2 AGN absorbs
much of the energy of optical and ultraviolet photons. This dust
emits at mid-infrared wavelengths, to which the dust is more
transparent (though obscuration may be significant even at these
wavelengths; \citealt{nenk08}). Thus observations of type 2 quasars in the
mid-infrared may provide a more direct probe of their bolometric
emission than the narrow emission lines we have used so far. We
matched our sample against the AllWISE \citep{wrig10, cutr13a} catalog
of the Wide-field Infrared Survey Explorer (WISE) using a
5\arcsec\ matching radius, picking the nearest match when multiple
matches are found. Approximately 97 per cent of the objects in our type 2
quasar sample have a successful match in AllWISE. To estimate the
contamination rate, we offset the positions of our sources by
1\arcmin\ and re-match, resulting in a matching rate of 9 per cent within
5\arcsec. Therefore, a few per cent of our AllWISE matches may be
random associations, or have their fluxes contaminated by unrelated
objects.   
 
\begin{figure}[htb]
\includegraphics[scale=0.40]{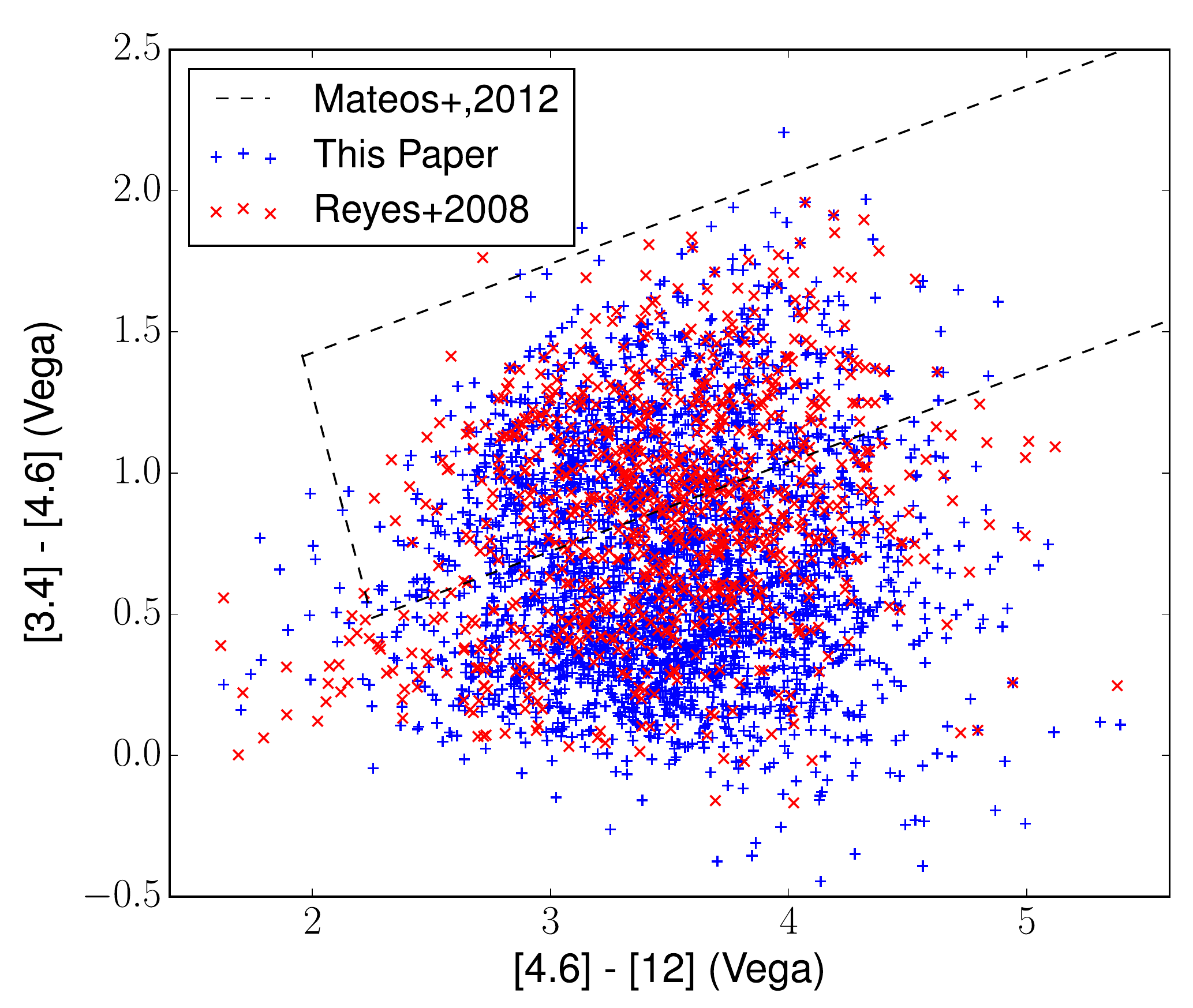}
\caption{The distribution of our sample and that of \citet{reye08} in the WISE ($[3.4] - [4.6]$ vs $[4.6] - [12]$) colors. The `wedge" denoted by the dashed lines is the luminous AGN selection region as defined by \citet{2012Mateos, mate13}. Only 34 per cent of the BOSS type 2 quasars are within this region, indicating that many type 2 quasars would be missed in infrared-selected samples.\label{fig:w1w2w3}}
\end{figure}

We fit piece-wise power-laws between each pair of adjacent WISE bands to determine a flux density, and thus a luminosity ($\nu L_\nu$) at rest frame 5 and 12 $\mu$m.   
  Figure~\ref{fig:w1w2w3} shows the distribution of our sources in
  WISE color space, using the filters at 3.4, 4.6, and 12$\mu$m. The
  median SNRs of the detections of our
  objects in the 3.4, 4.6, 12 and 22 $\mu$m bands are 26.7, 16.9, 6.8, 
  and 4.0 respectively. The figure also includes
  type 2 quasars from \citet{reye08}; the two distributions are quite
  similar.  The ``wedge" denoted by the dashed lines is the luminous
  AGN selection region as defined by \citet{2012Mateos, mate13},
  analogous to other proposed mid-infrared color cuts used for
  obscured AGN selection \citep{lacy04, ster05, ster12b}. It is
  striking that only 34 per cent of our sample is encompassed by this
  wedge; thus, there is a substantial number of sources with strong
  optical signatures of a type 2 quasar which would not be identified
  by the standard color-based infrared selection criteria.  

In Vega magnitudes, $[3.4]-[4.6]\simeq 0$ corresponds to the color of
an old stellar population dominated by the Rayleigh-Jeans tail of the
spectral energy distribution of stellar photospheres, thus there are
few objects bluer than this cutoff. In the absence of any thermal
re-emission by dust, this would also be the typical color of a type 2
quasar whose mid-infrared emission is dominated by the host
galaxy. Contribution of warm dust emission moves sources to the right
(toward the redder $[3.4]-[12]$ color) and contribution of hot dust
emission moves sources upward (toward the redder $[3.4]-[4.6]$
color). Type 2 quasars can be obscured even at mid-infrared
wavelengths, which is known both from theoretical models
\citep{pier92} and observations which show that they are significantly
redder in the infrared than type 1 quasars \citep{liu13b}. Therefore,
the hot dust contribution is not strong enough in more than half of
the sample to push the objects into the mid-infrared wedge. It
    has been suggested that the objects that lie outside the wedge
    have WISE fluxes dominated by starlight
    \citep{alon06,ecka10,donl12,mate13}, but that would be surprising given
    the high inferred AGN luminosities for our sample.

\subsection{\oiii\ properties and bolometric luminosity}
\label{subsec:OIIIvsIR}

\begin{figure}[htb]
\centering
\includegraphics[scale=0.42]{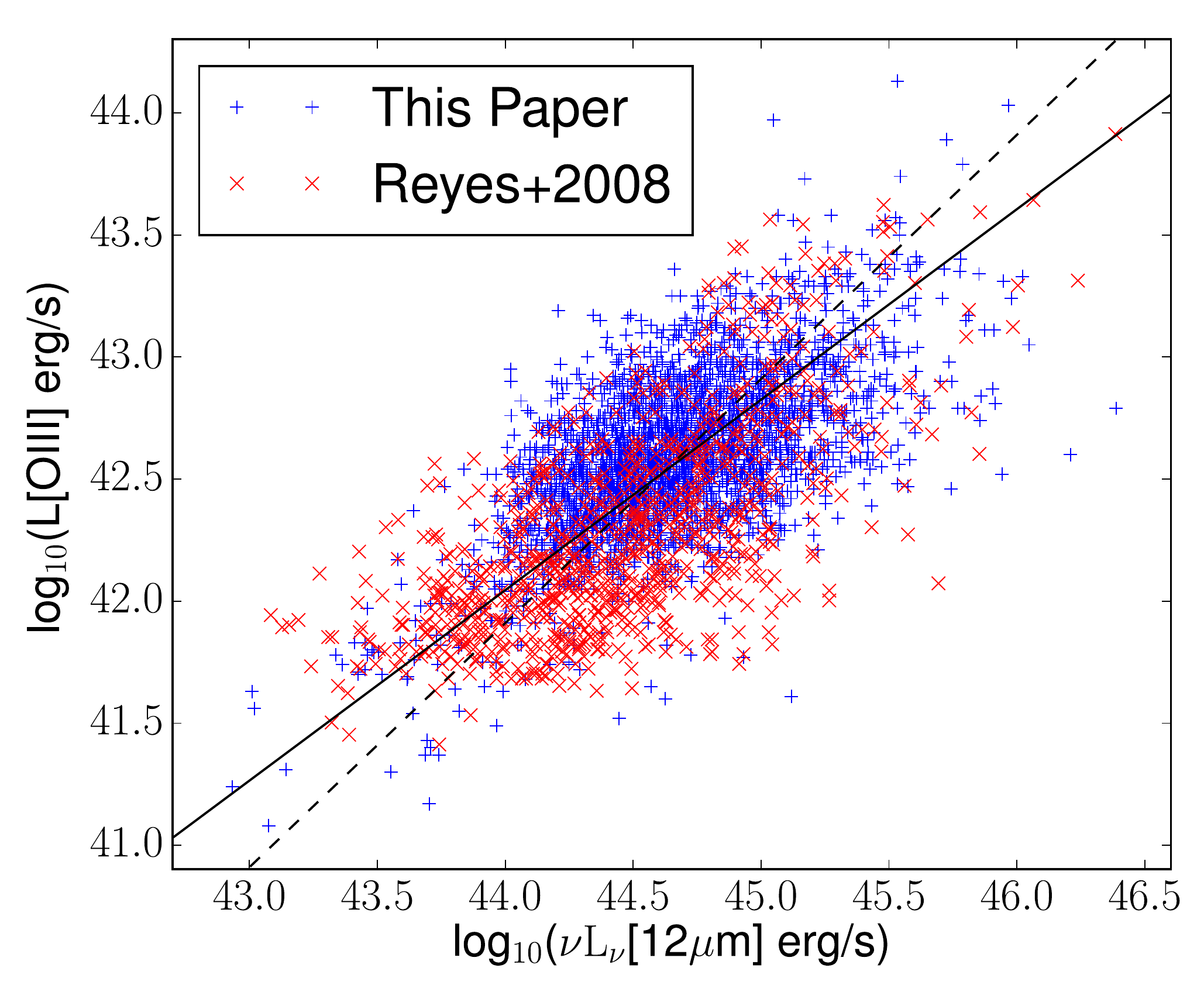}
\caption{The relationship between \oiii\ and rest-frame 12$\mu$m
  luminosity for the quasars in our sample and those of
  \citet{reye08}. All objects with WISE counterparts are shown
    here, not just those lying within the wedge of
    Figure~\ref{fig:w1w2w3}. The solid line is the best-fit power-law
  obtained by minimizing perpendicular residuals and the dashed line
  is the best-fit linear dependence, with best fits quantified by
  equations (\ref{eq_fits}).} 
\label{fig:oiii_ir}
\end{figure}

Figure~\ref{fig:oiii_ir} shows the relationship between the \oiii\ luminosities (based on the model fits described in
Section~\ref{subsec:OIII}) with the 
rest-frame 12$\mu$m luminosity for all objects with WISE
  counterparts in our sample and that of \citet{reye08}. The 12$\mu$m
luminosity is a proxy for a bolometric
luminosity, or at least the luminosity associated with hot dust close to the
central engine. The two luminosities are strongly correlated, suggesting that \oiii\ is a
useful, albeit rough, proxy for bolometric luminosity in obscured AGN
\citep{heck04}.  We calculate the best fits to the joint sample using
two methods: (i) we make the least-squares fit by minimizing
perpendicular offsets and (ii) we calculate the best-fit linear
relationship (i.e., with slope equal to unity).  The resulting best fits are: 

\begin{align}
& \log_{10}\left(\frac{\rm L_{[OIII]}}{\rm erg\ s^{-1}}\right) - 42.5 =  (0.78\pm 0.04)
\times \\ 
& \left[\log_{10}\left(\frac{\rm \nu L_{\nu}[12\mu m]}{\rm erg\ s^{-1}}\right)-44.5\right]
+ (-0.065\pm 0.017); \nonumber \\
& \log_{10}\left(\frac{\rm L_{[OIII]}}{\rm erg\ s^{-1}}\right) =
\log_{10}\left(\frac{\rm \nu L_{\nu}[12\mu m]}{\rm erg\ s^{-1}}\right)
- (2.090\pm 0.005).
\label{eq_fits}
\end{align}

These should be considered approximate scaling relations, because both
our new sample and the \citet{reye08} sample are affected by their
respective \oiii\ luminosity cutoffs (visible in Figure
\ref{fig:oiii_ir}). Furthermore, there are fewer objects at high
luminosity, so that our fit is more heavily
weighted by the less luminous objects. The correlation between
\oiii\ luminosity and 12\micron\ luminosity is tighter than the one
between \oiii\ luminosity and 5\micron\ luminosity \citep{zaka14},
presumably because the 5\micron\ luminosity is more strongly affected
by geometric effects and dust extinction. 

 [O$\,${\scriptsize III}] kinematics, in turn, can be a useful proxy for the strength of
 the quasar-driven outflows on host galaxy scales. In low-luminosity
 AGNs, the kinematics of the forbidden emission lines are strongly
 correlated with galaxy rotation and/or bulge velocity dispersion
 \citep{wils85, gree05o3}, indicating that the emission-line gas is in
 dynamical equilibrium with the galaxy. This is not the case in
 quasars \citep{zaka14}, where the characteristic velocities probed by
 \oiii\ emission are too high to be contained by the galactic
 potential. The velocity width asymmetry and 
 kurtosis of \oiii\ are all correlated with one another, suggesting that any of
 these values can serve as a proxy for outflow
 strength. \citet{zaka14} found that the strongest correlations are
 between the velocity width of the \oiii\ line (as measured by the
 $w_{90}$ parameter, Section~\ref{subsec:OIII}), and radio and
 infrared emission in the \citet{reye08} sample.  

\begin{figure}[t]
\centering
\includegraphics[width=9cm]{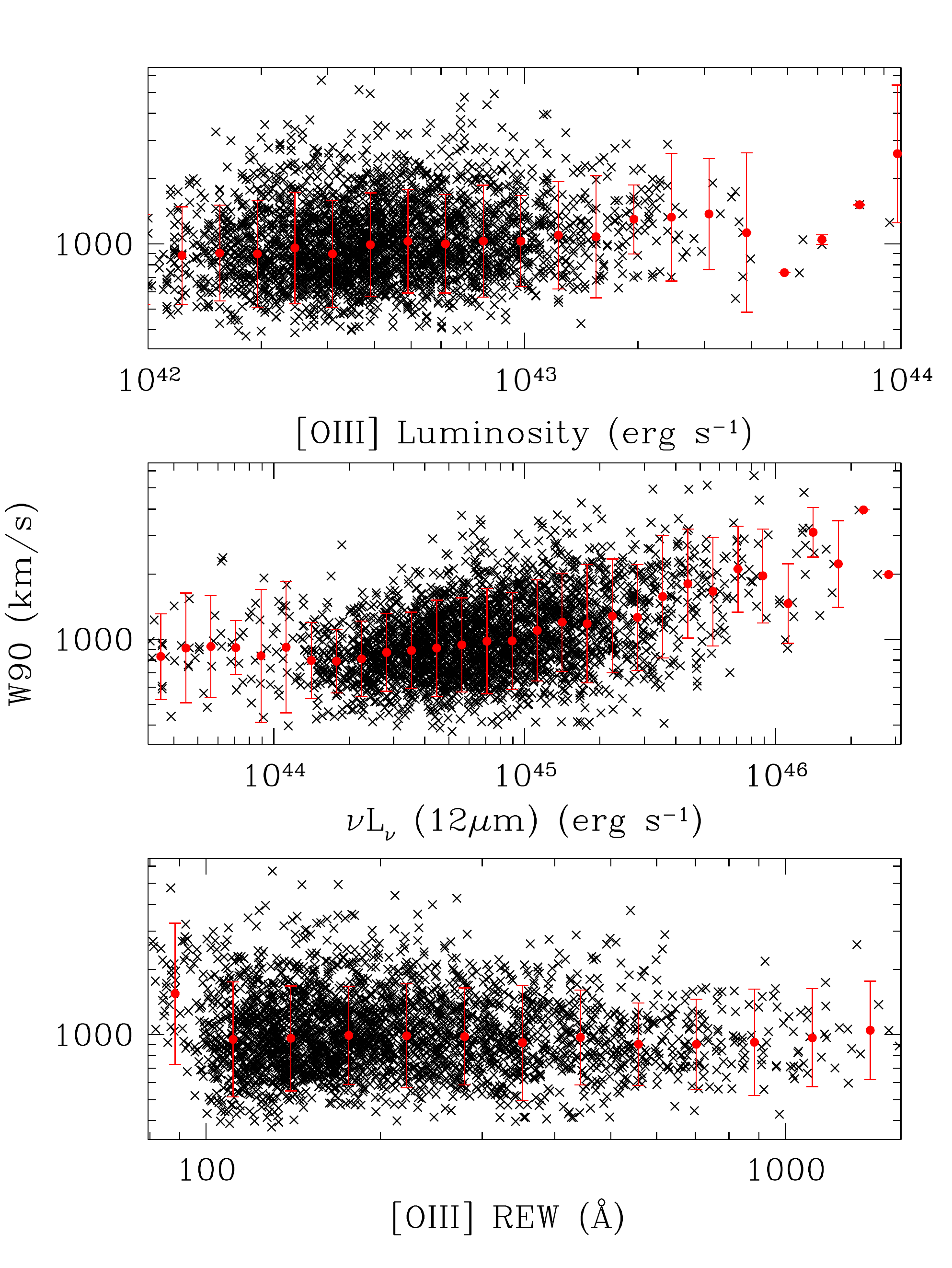}
\caption{The relationships between $w_{90}$ (the velocity width of
  \oiii\ containing 90 per cent of line power) and \oiii\ line luminosity,
  12$\mu$m infrared luminosity, and \oiii\ rest equivalent width. In each
  panel, the red points are median values in bins, with the
  interquartile range indicated. \oiii\ velocity width, which is a
  proxy for quasar wind activity, is most strongly correlated with
  quasar infrared luminosity.  
\label{fig:lum_width}}\end{figure}

In Figure~\ref{fig:lum_width} we investigate the relationship between
$w_{90}$, the velocity width containing 90 per cent of line power, with the
\oiii\ luminosity, the rest-frame 12$\mu$m luminosity, and the
\oiii\ equivalent width for the objects in our new BOSS 
sample. There is no correlation with equivalent width, and a weak one
with \oiii\ luminosity.  The correlation with infrared luminosity,
however, is quite strong, suggesting that indeed the velocity width
reflects the outflow velocity and that the outflow activity is driven
by the bolometric luminosity of the AGN. \citet{zaka16b} have found
objects at the peak epoch of quasar activity at $z \sim 2.5$ that lie 
at the extreme end of this diagram -- with extremely high infrared
luminosities and extremely broad \oiii; $w_{90}$ up to 5000 km
s$^{-1}$. 

\subsection{Extreme \NeV\ emitters}

Our sample includes objects with very strong
\NeV$\lambda$3426\AA\ emission, including sources with
\NeV/H$\beta>1$. Among type 2 quasars in the \citet{reye08} sample,
the mean and standard deviation of the quantity $\log$(\NeV/H$\beta$)
are $-0.3$ and 0.2, respectively, with only 6 per cent of objects showing
\NeV/H$\beta>1$. In our newly selected BOSS sample, these values are,
correspondingly, $-0.3$ dex, 0.3 dex and 13 per cent. The higher
fraction of objects with \NeV/H$\beta>$1 is likely due to our explicit
selection requirement that \NeV\ be detected in the $z>0.52$
subsample.  

\begin{figure}
\centering
\includegraphics[scale=0.43]{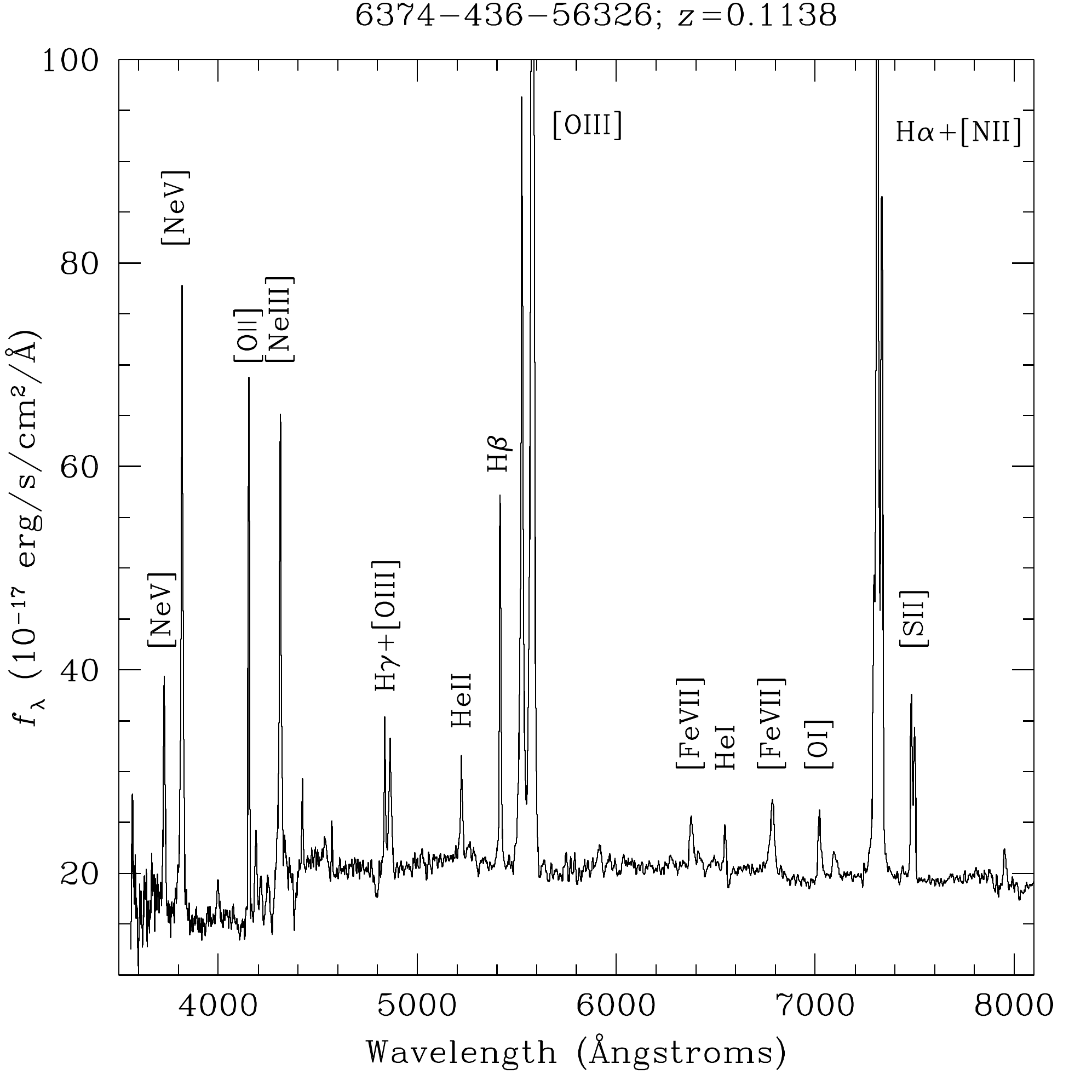} 
\caption{The BOSS spectrum (smoothed with a 5-pixel boxcar) of an
  extreme \NeV\ emitter from the newly selected BOSS sample.
  Prominent emission lines are marked. The \oiii\ and H$\alpha$ line
  peaks are offscale.  The stellar continuum is also apparent, showing
 Calcium K absorption and a strong Balmer break.}
  \label{fig:ext}
\end{figure}

One example of such an object from the BOSS sample is shown in Figure
\ref{fig:ext}. In addition to the very strong
\NeV\ (\NeV/H$\beta\simeq 2$ in this source), these objects also show
unusual emission 
features at 5721\AA\ and 6087\AA, which we identify as transitions of
[Fe{\scriptsize VII}] \citep{rose15a}. The ionization potentials of
\NeV\ and [Fe{\scriptsize VII}] are 97 and 99 eV, respectively, almost
identical, and thus it is not surprising that the strength of these
features are strongly correlated. 

\citet{rose15a,rose15b} deem such objects `coronal line forest AGNs'
and explore the hypothesis that these emission lines arise from the
inner wall of the obscuring material. It is thus rather unusual to see
these features in type 2 quasars, where it is expected that the line of
sight to this
emitting region should be obscured. \citet{rose15b} argue that even in
the classical unification model with a toroidal obscuring region, a
small fraction of viewing directions might result in both strong
coronal lines and obscured broad-line region (another such example from the
\citealt{reye08} sample is analyzed by \citealt{vill11a}). This
picture is consistent with the distribution of dust temperatures in
these objects and in type 2 quasars, in that coronal-line AGNs are warmer as
inferred from the infrared colors than are other type 2 quasars. The statistics of the coronal-line
AGNs in the type 2 population might offer clues to the geometric
structure of the obscuring material and provide constraints on its
clumpiness \citep{nenk02}. 

\subsection{Extended emission-line region}

\begin{figure}[t]
\centering
\includegraphics[width=8.5cm]{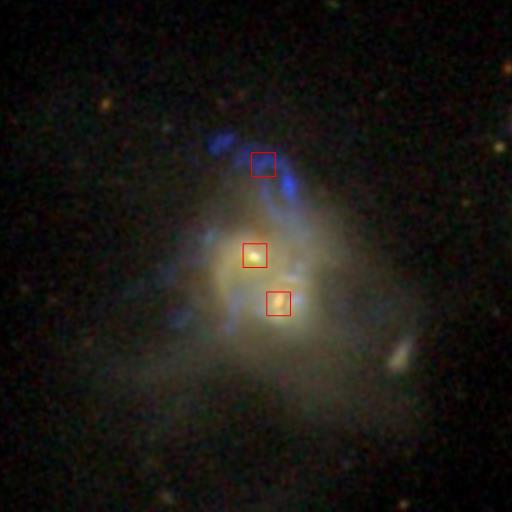}
\caption{The SDSS image of Mrk 266. The image is $51.2''$ on a
  side; North is up and East is to the left.  This is a $gri$ color
  composite, prepared following the approach of \citet{lupt04}.  The
  squares indicate the regions which have SDSS spectra; the SDSS image
  deblender identified the blue region in in the north as a separate
  ``galaxy'' . The northernmost spectrum was identified using our
  spectroscopic search for objects with high equivalent width, high
  ionization emission lines. The blue color of the nebular emission  in this
  region is due to the strong \oiii\ dominating the $g-$band emission in
  this extended nebula. 
\label{fig:voorwerp_image}}\end{figure}

One intriguing object identified using our initial selection is the
spectrum of an extended emission-line region associated with the
merging pair Mrk 
266 (Fig.~\ref{fig:voorwerp_image}), photo-ionized by the AGN in one of the merging nuclei. This
region of ionized gas, roughly 12 kpc from the central nucleus
\citep{hutc88,ishi00,mazz12}, has essentially no associated starlight,
and thus displays \oiii\ with a very high equivalent width. Another
well-known example of such an extended emission line nebula with AGN
line ratios is Hanny's
Voorwerp \citep{lint09}, which is found near a galaxy which no longer
hosts an active nucleus but presumably did in the recent past. A
systematic search for such objects in the SDSS data is conducted by
Sun et al. (in prep.). We removed the off-nuclear spectrum of Mrk 266
from the sample because it is not an integrated spectrum of the entire
galaxy.  

\subsection{Composite spectrum}

\begin{figure}[t]
\centering
\includegraphics[width=9cm]{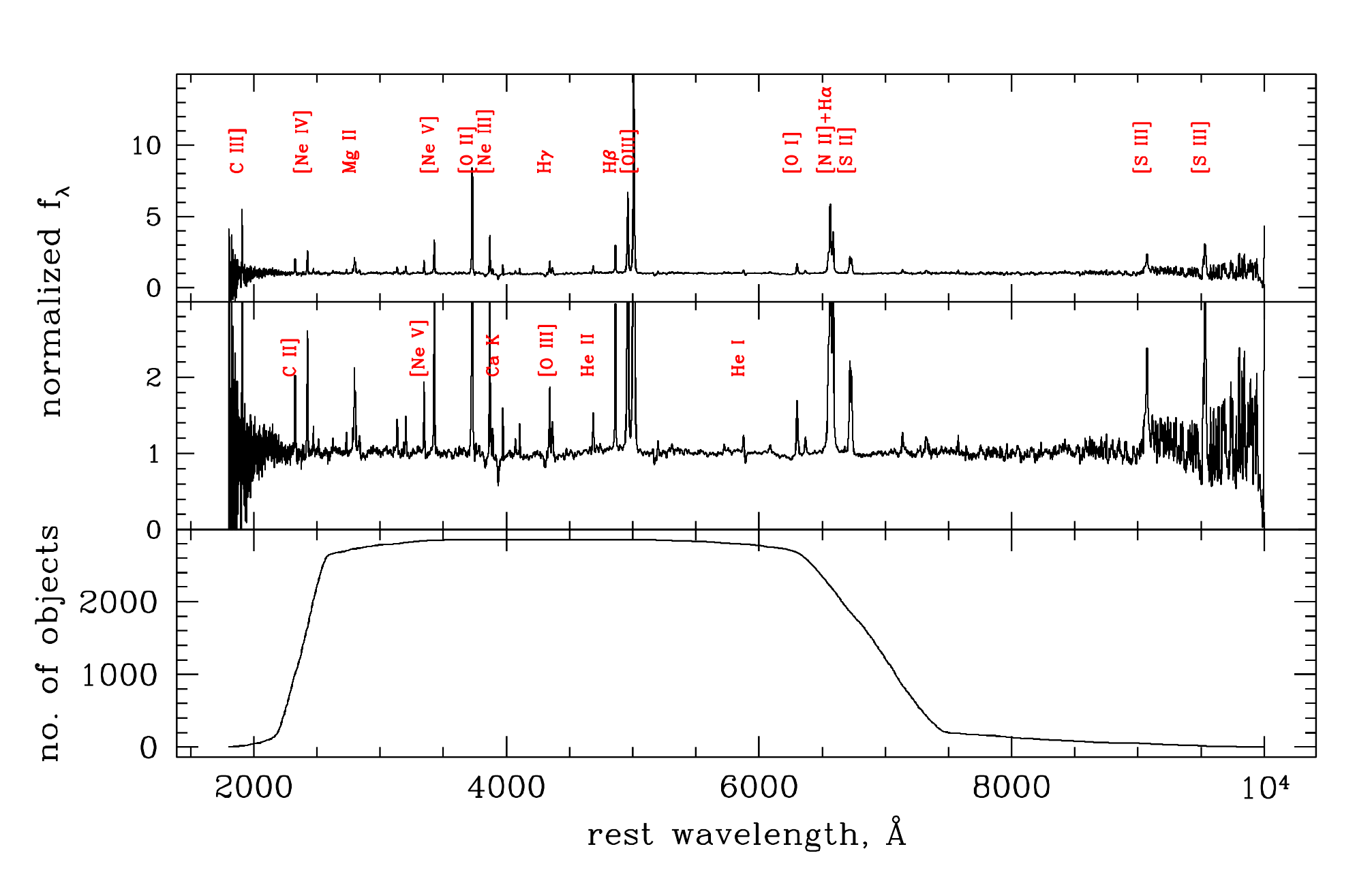}
\caption{The composite spectrum of all BOSS type 2 quasars normalized by an approximate continuum obtained by spline-interpolating between relatively line-free regions. The top spectrum shows the brightest emission lines, the middle panel shows a zoom in on the same data, and the bottom panel shows the number of objects contributing at each wavelength. Only a small subset of all detected lines are marked, following line identifications by \citet{vand01}. \label{fig:composite}}
\end{figure}

We construct a composite spectrum from our sample by shifting all spectra to their rest frames (using the adopted redshifts listed in the catalog), rebinning the spectra onto a common rest wavelength grid and calculating the error-weighted average. In Figure \ref{fig:composite} we further normalize the composite spectrum by a continuum obtained by spline-interpolating between relatively line-free regions. In this presentation the overall continuum shape is lost, but the equivalent widths of the features are preserved. We provide the composite spectrum in an online FITS table (\url{http://zakamska.johnshopkins.edu/data.htm}). The high SNR continuum reveals a multitude of emission features due to the quasar-ionized gas and absorption features due to the stellar photospheres and interstellar medium of the host galaxies. 

\section{Conclusions}
\label{sec:conc}

In this paper we present a sample of 2758 type 2 (obscured) quasars at $z\la 1$ selected from the SDSS-III/BOSS spectroscopic database. We aim to select sources with high equivalent width emission lines, REW(\oiii$\lambda$5007\AA)$>$100\AA. At low redshifts ($z<0.52$) we use standard emission-line diagnostic diagrams to separate type 2 candidates from star-forming galaxies. At higher redshifts ($z>0.52$), when H$\alpha$ and other diagnostics move out of the BOSS spectral coverage, we require a detection of \NeV$\lambda$3426\AA\ which requires ionization by an AGN and narrow H$\beta$ to separate type 2 quasars from type 1 quasars. 

An interesting subsample of 508 objects has erroneous or uncertain redshifts in the SDSS database. We select such sources by assuming that \oiii\ -- one of the strongest lines in the type 2 quasar spectra -- is mistaken for another strong line, either Ly$\alpha$ or H$\alpha$, by the pipeline. Additionally, we examine sources with high REW of \oiii\ and with redshifts flagged by the SDSS database pipeline as uncertain. These sources tend to have strongly kinematically disturbed emission lines and they are therefore poorly matched against the standard templates. Because our selection algorithms for such sources are not exhaustive, small numbers of interesting type 2 quasars with broad \oiii\ could remain unidentified in the SDSS spectroscopic database.

While in low-luminosity AGNs the kinematics of the forbidden emission lines tend to trace the potential of the AGN host galaxy, in powerful quasars \oiii\ profile shapes appear to be strongly related to quasar-driven winds. We conduct multi-Gaussian decomposition of \oiii\ for all objects in this sample and calculate all commonly used non-parametric measures of \oiii\ profile shape. We release complete kinematic decomposition information for both the new catalog of BOSS type 2 quasars and for our previous kinematic analysis of 568 luminous ($L$\oiii$>1.2\times 10^{42}\,\ergs$) type 2 quasars selected from SDSS I/II \citep{reye08, zaka14}. We also identify 654 candidate objects with double-peaked \oiii\ profiles which could be interesting for studies of quasar-driven winds or of binary AGNs. 

As we determine by matching the sample to the WISE survey, the type 2
quasars presented here have high infrared luminosities, with median
$\nu L_{\nu}=1.7\times 10^{44}\,\ergs$ at rest-frame
5\micron\ and $4.2\times 10^{44}\,\ergs$ at rest-frame
12\micron. If quasars were isotropic emitters at 12\micron, we could
apply a typical bolometric correction at this wavelength of $\sim 9$
\citep{rich06} to estimate the median bolometric luminosity of our
sample to be $\sim 4\times 10^{45}\,\ergs$. Intriguingly,
despite very high luminosities, fewer than half of the objects in our
sample have $[3.6-4.5]$ color red enough for them to be selected using
the common infrared color selection methods used to identify
AGNs. Thus it is likely that type 2 quasars are obscured even at
mid-infrared wavelengths, so that hot dust emission from the inner
parts of the obscuring material remains largely invisible to the
observer \citep{liu13b}. Therefore, the bolometric corrections are
likely higher than those for type 1 quasars, and so then is our
estimated median bolometric luminosity.  

The demographics of quasars remain an interesting unsolved problem in
astronomy, with obscured quasars now thought to play an important role
in galaxy evolution. This work, alongside other approaches, makes it
clear that different selection methods result in largely different
samples of objects. Samples selected by infrared, optical and X-ray
methods overlap at the tens of per cent level \citep{lacy13}, but none
of the selection methods results in a complete sample. The combination of
multi-wavelength approaches and extensive studies of samples selected
at different wavelengths will be required to measure the demographics
of quasars and to determine the geometry and the spatial structure of
the obscuring material.  

\acknowledgments

Funding for SDSS-III has been provided by the Alfred P. Sloan
Foundation, the Participating Institutions, the National Science
Foundation, and the U.S. Department of Energy Office of Science. The
SDSS-III web site is \url{http://www.sdss3.org/}. 

SDSS-III is managed by the Astrophysical Research Consortium for the
Participating Institutions of the SDSS-III Collaboration including the
University of Arizona, the Brazilian Participation Group, Brookhaven
National Laboratory, Carnegie Mellon University, University of
Florida, the French Participation Group, the German Participation
Group, Harvard University, the Instituto de Astrofisica de Canarias,
the Michigan State/Notre Dame/JINA Participation Group, Johns Hopkins
University, Lawrence Berkeley National Laboratory, Max Planck
Institute for Astrophysics, Max Planck Institute for Extraterrestrial
Physics, New Mexico State University, New York University, Ohio State
University, Pennsylvania State University, University of Portsmouth,
Princeton University, the Spanish Participation Group, University of
Tokyo, University of Utah, Vanderbilt University, University of
Virginia, University of Washington, and Yale University. 

\bibliographystyle{aasjournal}
\bibliography{master_0612,additional}

\begin{thebibliography}{}
\expandafter\ifx\csname natexlab\endcsname\relax\def\natexlab#1{#1}\fi

\bibitem[{{Adelman-McCarthy} {et~al.}(2008){Adelman-McCarthy}, {Ag{\"u}eros},
  {Allam}, {Allende Prieto}, {Anderson}, {Anderson}, {Annis}, {Bahcall},
  {Bailer-Jones}, {Baldry}, {Barentine}, {Bassett}, {Becker}, {Beers}, {Bell},
  {Berlind}, {Bernardi}, {Blanton}, {Bochanski}, {Boroski}, {Brinchmann},
  {Brinkmann}, {Brunner}, {Budav{\'a}ri}, {Carliles}, {Carr}, {Castander},
  {Cinabro}, {Cool}, {Covey}, {Csabai}, {Cunha}, {Davenport}, {Dilday}, {Doi},
  {Eisenstein}, {Evans}, {Fan}, {Finkbeiner}, {Friedman}, {Frieman},
  {Fukugita}, {G{\"a}nsicke}, {Gates}, {Gillespie}, {Glazebrook}, {Gray},
  {Grebel}, {Gunn}, {Gurbani}, {Hall}, {Harding}, {Harvanek}, {Hawley},
  {Hayes}, {Heckman}, {Hendry}, {Hindsley}, {Hirata}, {Hogan}, {Hogg}, {Hyde},
  {Ichikawa}, {Ivezi{\'c}}, {Jester}, {Johnson}, {Jorgensen}, {Juri{\'c}},
  {Kent}, {Kessler}, {Kleinman}, {Knapp}, {Kron}, {Krzesinski}, {Kuropatkin},
  {Lamb}, {Lampeitl}, {Lebedeva}, {Lee}, {Leger}, {L{\'e}pine}, {Lima}, {Lin},
  {Long}, {Loomis}, {Loveday}, {Lupton}, {Malanushenko}, {Malanushenko},
  {Mandelbaum}, {Margon}, {Marriner}, {Mart{\'{\i}}nez-Delgado}, {Matsubara},
  {McGehee}, {McKay}, {Meiksin}, {Morrison}, {Munn}, {Nakajima}, {Neilsen},
  {Newberg}, {Nichol}, {Nicinski}, {Nieto-Santisteban}, {Nitta}, {Okamura},
  {Owen}, {Oyaizu}, {Padmanabhan}, {Pan}, {Park}, {Peoples}, {Pier}, {Pope},
  {Purger}, {Raddick}, {Re Fiorentin}, {Richards}, {Richmond}, {Riess}, {Rix},
  {Rockosi}, {Sako}, {Schlegel}, {Schneider}, {Schreiber}, {Schwope}, {Seljak},
  {Sesar}, {Sheldon}, {Shimasaku}, {Sivarani}, {Smith}, {Snedden}, {Steinmetz},
  {Strauss}, {SubbaRao}, {Suto}, {Szalay}, {Szapudi}, {Szkody}, {Tegmark},
  {Thakar}, {Tremonti}, {Tucker}, {Uomoto}, {Vanden Berk}, {Vandenberg},
  {Vidrih}, {Vogeley}, {Voges}, {Vogt}, {Wadadekar}, {Weinberg}, {West},
  {White}, {Wilhite}, {Yanny}, {Yocum}, {York}, {Zehavi}, \& {Zucker}}]{adel08}
{Adelman-McCarthy}, J.~K., {Ag{\"u}eros}, M.~A., {Allam}, S.~S., {et~al.} 2008,
  \apjs, 175, 297

\bibitem[{{Ahn} {et~al.}(2012){Ahn}, {Alexandroff}, {Allende Prieto},
  {Anderson}, {Anderton}, {Andrews}, {Aubourg}, {Bailey}, {Balbinot}, {Barnes},
  \& et~al.}]{DR9}
{Ahn}, C.~P., {Alexandroff}, R., {Allende Prieto}, C., {et~al.} 2012, \apjs,
  203, 21

\bibitem[{{Aihara} {et~al.}(2011){Aihara}, {Allende Prieto}, {An}, {Anderson},
  {Aubourg}, {Balbinot}, {Beers}, {Berlind}, {Bickerton}, {Bizyaev}, {Blanton},
  {Bochanski}, {Bolton}, {Bovy}, {Brandt}, {Brinkmann}, {Brown}, {Brownstein},
  {Busca}, {Campbell}, {Carr}, {Chen}, {Chiappini}, {Comparat}, {Connolly},
  {Cortes}, {Croft}, {Cuesta}, {da Costa}, {Davenport}, {Dawson}, {Dhital},
  {Ealet}, {Ebelke}, {Edmondson}, {Eisenstein}, {Escoffier}, {Esposito},
  {Evans}, {Fan}, {Femen{\'{\i}}a Castell{\'a}}, {Font-Ribera}, {Frinchaboy},
  {Ge}, {Gillespie}, {Gilmore}, {Gonz{\'a}lez Hern{\'a}ndez}, {Gott}, {Gould},
  {Grebel}, {Gunn}, {Hamilton}, {Harding}, {Harris}, {Hawley}, {Hearty}, {Ho},
  {Hogg}, {Holtzman}, {Honscheid}, {Inada}, {Ivans}, {Jiang}, {Johnson},
  {Jordan}, {Jordan}, {Kazin}, {Kirkby}, {Klaene}, {Knapp}, {Kneib},
  {Kochanek}, {Koesterke}, {Kollmeier}, {Kron}, {Lampeitl}, {Lang}, {Le Goff},
  {Lee}, {Lin}, {Long}, {Loomis}, {Lucatello}, {Lundgren}, {Lupton}, {Ma},
  {MacDonald}, {Mahadevan}, {Maia}, {Makler}, {Malanushenko}, {Malanushenko},
  {Mandelbaum}, {Maraston}, {Margala}, {Masters}, {McBride}, {McGehee},
  {McGreer}, {M{\'e}nard}, {Miralda-Escud{\'e}}, {Morrison}, {Mullally},
  {Muna}, {Munn}, {Murayama}, {Myers}, {Naugle}, {Neto}, {Nguyen}, {Nichol},
  {O'Connell}, {Ogando}, {Olmstead}, {Oravetz}, {Padmanabhan},
  {Palanque-Delabrouille}, {Pan}, {Pandey}, {P{\^a}ris}, {Percival},
  {Petitjean}, {Pfaffenberger}, {Pforr}, {Phleps}, {Pichon}, {Pieri}, {Prada},
  {Price-Whelan}, {Raddick}, {Ramos}, {Reyl{\'e}}, {Rich}, {Richards}, {Rix},
  {Robin}, {Rocha-Pinto}, {Rockosi}, {Roe}, {Rollinde}, {Ross}, {Ross},
  {Rossetto}, {S{\'a}nchez}, {Sayres}, {Schlegel}, {Schlesinger}, {Schmidt},
  {Schneider}, {Sheldon}, {Shu}, {Simmerer}, {Simmons}, {Sivarani}, {Snedden},
  {Sobeck}, {Steinmetz}, {Strauss}, {Szalay}, {Tanaka}, {Thakar}, {Thomas},
  {Tinker}, {Tofflemire}, {Tojeiro}, {Tremonti}, {Vandenberg}, {Vargas
  Maga{\~n}a}, {Verde}, {Vogt}, {Wake}, {Wang}, {Weaver}, {Weinberg}, {White},
  {White}, {Yanny}, {Yasuda}, {Yeche}, \& {Zehavi}}]{DR8}
{Aihara}, H., {Allende Prieto}, C., {An}, D., {et~al.} 2011, \apjs, 193, 29

\bibitem[{{Alam} {et~al.}(2015){Alam}, {Albareti}, {Allende Prieto}, {Anders},
  {Anderson}, {Anderton}, {Andrews}, {Armengaud}, {Aubourg}, {Bailey}, \&
  et~al.}]{alam15}
{Alam}, S., {Albareti}, F.~D., {Allende Prieto}, C., {et~al.} 2015, \apjs, 219,
  12

\bibitem[{{Alexandroff} {et~al.}(2013){Alexandroff}, {Strauss}, {Greene},
  {Zakamska}, {Ross}, {Brandt}, {Liu}, {Smith}, {Ge}, {Hamann}, {Myers},
  {Petitjean}, {Schneider}, {Yesuf}, \& {York}}]{alex13}
{Alexandroff}, R., {Strauss}, M.~A., {Greene}, J.~E., {et~al.} 2013, \mnras,
  435, 3306

\bibitem[{{Alonso-Herrero} {et~al.}(2006){Alonso-Herrero},
  {P{\'e}rez-Gonz{\'a}lez}, {Alexander}, {Rieke}, {Rigopoulou}, {Le Floc'h},
  {Barmby}, {Papovich}, {Rigby}, {Bauer}, {Brandt}, {Egami}, {Willner}, {Dole},
  \& {Huang}}]{alon06}
{Alonso-Herrero}, A., {P{\'e}rez-Gonz{\'a}lez}, P.~G., {Alexander}, D.~M.,
  {et~al.} 2006, \apj, 640, 167

\bibitem[{{Antonucci}(1993)}]{anto93}
{Antonucci}, R. 1993, \araa, 31, 473

\bibitem[{{Antonucci} \& {Miller}(1985)}]{anto85}
{Antonucci}, R.~R.~J., \& {Miller}, J.~S. 1985, \apj, 297, 621

\bibitem[{{Aubourg} {et~al.}(2015){Aubourg}, {Bailey}, {Bautista}, {Beutler},
  {Bhardwaj}, {Bizyaev}, {Blanton}, {Blomqvist}, {Bolton}, {Bovy},
  {Brewington}, {Brinkmann}, {Brownstein}, {Burden}, {Busca}, {Carithers},
  {Chuang}, {Comparat}, {Croft}, {Cuesta}, {Dawson}, {Delubac}, {Eisenstein},
  {Font-Ribera}, {Ge}, {Le Goff}, {Gontcho}, {Gott}, {Gunn}, {Guo}, {Guy},
  {Hamilton}, {Ho}, {Honscheid}, {Howlett}, {Kirkby}, {Kitaura}, {Kneib},
  {Lee}, {Long}, {Lupton}, {Maga{\~n}a}, {Malanushenko}, {Malanushenko},
  {Manera}, {Maraston}, {Margala}, {McBride}, {Miralda-Escud{\'e}}, {Myers},
  {Nichol}, {Noterdaeme}, {Nuza}, {Olmstead}, {Oravetz}, {P{\^a}ris},
  {Padmanabhan}, {Palanque-Delabrouille}, {Pan}, {Pellejero-Ibanez},
  {Percival}, {Petitjean}, {Pieri}, {Prada}, {Reid}, {Rich}, {Roe}, {Ross},
  {Ross}, {Rossi}, {Rubi{\~n}o-Mart{\'{\i}}n}, {S{\'a}nchez}, {Samushia},
  {Santos}, {Sc{\'o}ccola}, {Schlegel}, {Schneider}, {Seo}, {Sheldon},
  {Simmons}, {Skibba}, {Slosar}, {Strauss}, {Thomas}, {Tinker}, {Tojeiro},
  {Vazquez}, {Viel}, {Wake}, {Weaver}, {Weinberg}, {Wood-Vasey}, {Y{\`e}che},
  {Zehavi}, {Zhao}, \& {BOSS Collaboration}}]{aubo15}
{Aubourg}, {\'E}., {Bailey}, S., {Bautista}, J.~E., {et~al.} 2015, \prd, 92,
  123516

\bibitem[{{Baldwin} {et~al.}(1981){Baldwin}, {Phillips}, \&
  {Terlevich}}]{bald81}
{Baldwin}, J.~A., {Phillips}, M.~M., \& {Terlevich}, R. 1981, \pasp, 93, 5

\bibitem[{{Bolton} {et~al.}(2012){Bolton}, {Schlegel}, {Aubourg}, {Bailey},
  {Bhardwaj}, {Brownstein}, {Burles}, {Chen}, {Dawson}, {Eisenstein}, {Gunn},
  {Knapp}, {Loomis}, {Lupton}, {Maraston}, {Muna}, {Myers}, {Olmstead},
  {Padmanabhan}, {P{\^a}ris}, {Percival}, {Petitjean}, {Rockosi}, {Ross},
  {Schneider}, {Shu}, {Strauss}, {Thomas}, {Tremonti}, {Wake}, {Weaver}, \&
  {Wood-Vasey}}]{bolt12}
{Bolton}, A.~S., {Schlegel}, D.~J., {Aubourg}, {\'E}., {et~al.} 2012, \aj, 144,
  144

\bibitem[{{Brandt} \& {Hasinger}(2005)}]{bran05}
{Brandt}, W.~N., \& {Hasinger}, G. 2005, \araa, 43, 827

\bibitem[{{Brusa} {et~al.}(2010){Brusa}, {Civano}, {Comastri}, {Miyaji},
  {Salvato}, {Zamorani}, {Cappelluti}, {Fiore}, {Hasinger}, {Mainieri},
  {Merloni}, {Bongiorno}, {Capak}, {Elvis}, {Gilli}, {Hao}, {Jahnke},
  {Koekemoer}, {Ilbert}, {Le Floc'h}, {Lusso}, {Mignoli}, {Schinnerer},
  {Silverman}, {Treister}, {Trump}, {Vignali}, {Zamojski}, {Aldcroft},
  {Aussel}, {Bardelli}, {Bolzonella}, {Cappi}, {Caputi}, {Contini},
  {Finoguenov}, {Fruscione}, {Garilli}, {Impey}, {Iovino}, {Iwasawa},
  {Kampczyk}, {Kartaltepe}, {Kneib}, {Knobel}, {Kovac}, {Lamareille},
  {Leborgne}, {Le Brun}, {Le Fevre}, {Lilly}, {Maier}, {McCracken}, {Pello},
  {Peng}, {Perez-Montero}, {de Ravel}, {Sanders}, {Scodeggio}, {Scoville},
  {Tanaka}, {Taniguchi}, {Tasca}, {de la Torre}, {Tresse}, {Vergani}, \&
  {Zucca}}]{brus10}
{Brusa}, M., {Civano}, F., {Comastri}, A., {et~al.} 2010, \apj, 716, 348

\bibitem[{{Comerford} {et~al.}(2012){Comerford}, {Gerke}, {Stern}, {Cooper},
  {Weiner}, {Newman}, {Madsen}, \& {Barrows}}]{come12}
{Comerford}, J.~M., {Gerke}, B.~F., {Stern}, D., {et~al.} 2012, \apj, 753, 42

\bibitem[{{Constantin} \& {Shields}(2003)}]{cons03}
{Constantin}, A., \& {Shields}, J.~C. 2003, \pasp, 115, 592

\bibitem[{{Cutri} \& {et al.}(2013)}]{cutr13a}
{Cutri}, R.~M., \& {et al.} 2013, VizieR Online Data Catalog, 2328

\bibitem[{{Dawson} {et~al.}(2013){Dawson}, {Schlegel}, \& et~al.}]{daws13}
{Dawson}, K.~S., {Schlegel}, D.~J., \& et~al. 2013, \aj, 145, 10

\bibitem[{{De Robertis} \& {Osterbrock}(1984)}]{dero84}
{De Robertis}, M.~M., \& {Osterbrock}, D.~E. 1984, \apj, 286, 171

\bibitem[{{Donley} {et~al.}(2012){Donley}, {Koekemoer}, {Brusa}, {Capak},
  {Cardamone}, {Civano}, {Ilbert}, {Impey}, {Kartaltepe}, {Miyaji}, {Salvato},
  {Sanders}, {Trump}, \& {Zamorani}}]{donl12}
{Donley}, J.~L., {Koekemoer}, A.~M., {Brusa}, M., {et~al.} 2012, \apj, 748, 142

\bibitem[{{Eckart} {et~al.}(2010){Eckart}, {McGreer}, {Stern}, {Harrison}, \&
  {Helfand}}]{ecka10}
{Eckart}, M.~E., {McGreer}, I.~D., {Stern}, D., {Harrison}, F.~A., \&
  {Helfand}, D.~J. 2010, \apj, 708, 584

\bibitem[{{Eisenhardt} {et~al.}(2012){Eisenhardt}, {Wu}, {Tsai}, {Assef},
  {Benford}, {Blain}, {Bridge}, {Condon}, {Cushing}, {Cutri}, {Evans},
  {Gelino}, {Griffith}, {Grillmair}, {Jarrett}, {Lonsdale}, {Masci}, {Mason},
  {Petty}, {Sayers}, {Stanford}, {Stern}, {Wright}, \& {Yan}}]{eise12}
{Eisenhardt}, P.~R.~M., {Wu}, J., {Tsai}, C.-W., {et~al.} 2012, \apj, 755, 173

\bibitem[{{Eisenstein} {et~al.}(2011){Eisenstein}, {Weinberg}, {Agol},
  {Aihara}, {Allende Prieto}, {Anderson}, {Arns}, {Aubourg}, {Bailey},
  {Balbinot}, \& et~al.}]{eise11}
{Eisenstein}, D.~J., {Weinberg}, D.~H., {Agol}, E., {et~al.} 2011, \aj, 142, 72

\bibitem[{{Gilli} {et~al.}(2010){Gilli}, {Vignali}, {Mignoli}, {Iwasawa},
  {Comastri}, \& {Zamorani}}]{gill10}
{Gilli}, R., {Vignali}, C., {Mignoli}, M., {et~al.} 2010, \aap, 519, A92

\bibitem[{{Glikman} {et~al.}(2012){Glikman}, {Urrutia}, {Lacy}, {Djorgovski},
  {Mahabal}, {Myers}, {Ross}, {Petitjean}, {Ge}, {Schneider}, \&
  {York}}]{glik12}
{Glikman}, E., {Urrutia}, T., {Lacy}, M., {et~al.} 2012, \apj, 757, 51

\bibitem[{{Greene} \& {Ho}(2005)}]{gree05o3}
{Greene}, J.~E., \& {Ho}, L.~C. 2005, \apj, 627, 721

\bibitem[{{Greene} {et~al.}(2011){Greene}, {Zakamska}, {Ho}, \&
  {Barth}}]{gree11}
{Greene}, J.~E., {Zakamska}, N.~L., {Ho}, L.~C., \& {Barth}, A.~J. 2011, \apj,
  732, 9

\bibitem[{{Greene} {et~al.}(2009){Greene}, {Zakamska}, {Liu}, {Barth}, \&
  {Ho}}]{gree09}
{Greene}, J.~E., {Zakamska}, N.~L., {Liu}, X., {Barth}, A.~J., \& {Ho}, L.~C.
  2009, \apj, 702, 441

\bibitem[{{Greene} {et~al.}(2012){Greene}, {Zakamska}, \& {Smith}}]{gree12}
{Greene}, J.~E., {Zakamska}, N.~L., \& {Smith}, P.~S. 2012, \apj, 746, 86

\bibitem[{{Hainline} {et~al.}(2013){Hainline}, {Hickox}, {Greene}, {Myers}, \&
  {Zakamska}}]{hain13}
{Hainline}, K.~N., {Hickox}, R., {Greene}, J.~E., {Myers}, A.~D., \&
  {Zakamska}, N.~L. 2013, \apj, 774, 145

\bibitem[{{Hao} {et~al.}(2005{\natexlab{a}}){Hao}, {Strauss}, {Tremonti},
  {Schlegel}, {Heckman}, {Kauffmann}, {Blanton}, {Fan}, {Gunn}, {Hall},
  {Ivezi{\'c}}, {Knapp}, {Krolik}, {Lupton}, {Richards}, {Schneider},
  {Strateva}, {Zakamska}, {Brinkmann}, {Brunner}, \& {Szokoly}}]{hao05a}
{Hao}, L., {Strauss}, M.~A., {Tremonti}, C.~A., {et~al.} 2005{\natexlab{a}},
  \aj, 129, 1783

\bibitem[{{Hao} {et~al.}(2005{\natexlab{b}}){Hao}, {Strauss}, {Fan},
  {Tremonti}, {Schlegel}, {Heckman}, {Kauffmann}, {Blanton}, {Gunn}, {Hall},
  {Ivezi{\'c}}, {Knapp}, {Krolik}, {Lupton}, {Richards}, {Schneider},
  {Strateva}, {Zakamska}, {Brinkmann}, \& {Szokoly}}]{hao05b}
{Hao}, L., {Strauss}, M.~A., {Fan}, X., {et~al.} 2005{\natexlab{b}}, \aj, 129,
  1795

\bibitem[{{Harrison} {et~al.}(2015){Harrison}, {Thomson}, {Alexander}, {Bauer},
  {Edge}, {Hogan}, {Mullaney}, \& {Swinbank}}]{harr15}
{Harrison}, C.~M., {Thomson}, A.~P., {Alexander}, D.~M., {et~al.} 2015, \apj,
  800, 45

\bibitem[{{Hasinger}(2008)}]{hasi08}
{Hasinger}, G. 2008, \aap, 490, 905

\bibitem[{{Heckman} {et~al.}(2004){Heckman}, {Kauffmann}, {Brinchmann},
  {Charlot}, {Tremonti}, \& {White}}]{heck04}
{Heckman}, T.~M., {Kauffmann}, G., {Brinchmann}, J., {et~al.} 2004, \apj, 613,
  109

\bibitem[{{Heckman} {et~al.}(1981){Heckman}, {Miley}, {van Breugel}, \&
  {Butcher}}]{heck81}
{Heckman}, T.~M., {Miley}, G.~K., {van Breugel}, W.~J.~M., \& {Butcher}, H.~R.
  1981, \apj, 247, 403

\bibitem[{{Hopkins} {et~al.}(2006){Hopkins}, {Hernquist}, {Cox}, {Di Matteo},
  {Robertson}, \& {Springel}}]{hopk06}
{Hopkins}, P.~F., {Hernquist}, L., {Cox}, T.~J., {et~al.} 2006, \apjs, 163, 1

\bibitem[{{Hopkins} {et~al.}(2007){Hopkins}, {Richards}, \&
  {Hernquist}}]{hopk07}
{Hopkins}, P.~F., {Richards}, G.~T., \& {Hernquist}, L. 2007, \apj, 654, 731

\bibitem[{{Hutchings} {et~al.}(1988){Hutchings}, {Neff}, \& {van
  Gorkom}}]{hutc88}
{Hutchings}, J.~B., {Neff}, S.~G., \& {van Gorkom}, J.~H. 1988, \aj, 96, 1227

\bibitem[{{Ishigaki} {et~al.}(2000){Ishigaki}, {Yoshida}, {Aoki}, {Ohtani},
  {Sugai}, {Hayashi}, {Ozaki}, {Hattori}, \& {Ishii}}]{ishi00}
{Ishigaki}, T., {Yoshida}, M., {Aoki}, K., {et~al.} 2000, \pasj, 52, 185

\bibitem[{{Kauffmann} {et~al.}(2003)}]{kauf03a}
{Kauffmann}, G., {et~al.} 2003, \mnras, 346, 1055

\bibitem[{{Kewley} {et~al.}(2001){Kewley}, {Dopita}, {Sutherland}, {Heisler},
  \& {Trevena}}]{kewl01}
{Kewley}, L.~J., {Dopita}, M.~A., {Sutherland}, R.~S., {Heisler}, C.~A., \&
  {Trevena}, J. 2001, \apj, 556, 121

\bibitem[{{Khachikian} \& {Weedman}(1974)}]{khac74}
{Khachikian}, E.~Y., \& {Weedman}, D.~W. 1974, \apj, 192, 581

\bibitem[{{Kirhakos} {et~al.}(1999){Kirhakos}, {Bahcall}, {Schneider}, \&
  {Kristian}}]{kirhakos99}
{Kirhakos}, S., {Bahcall}, J.~N., {Schneider}, D.~P., \& {Kristian}, J. 1999,
  \apj, 520, 67

\bibitem[{{Lacy} {et~al.}(2015){Lacy}, {Ridgway}, {Sajina}, {Petric}, {Gates},
  {Urrutia}, \& {Storrie-Lombardi}}]{lacy15}
{Lacy}, M., {Ridgway}, S.~E., {Sajina}, A., {et~al.} 2015, \apj, 802, 102

\bibitem[{{Lacy} {et~al.}(2007){Lacy}, {Sajina}, {Petric}, {Seymour},
  {Canalizo}, {Ridgway}, {Armus}, \& {Storrie-Lombardi}}]{lacy07}
{Lacy}, M., {Sajina}, A., {Petric}, A.~O., {et~al.} 2007, \apjl, 669, L61

\bibitem[{{Lacy} {et~al.}(2004){Lacy}, {Storrie-Lombardi}, {Sajina},
  {Appleton}, {Armus}, {Chapman}, {Choi}, {Fadda}, {Fang}, {Frayer},
  {Heinrichsen}, {Helou}, {Im}, {Marleau}, {Masci}, {Shupe}, {Soifer},
  {Surace}, {Teplitz}, {Wilson}, \& {Yan}}]{lacy04}
{Lacy}, M., {Storrie-Lombardi}, L.~J., {Sajina}, A., {et~al.} 2004, \apjs, 154,
  166

\bibitem[{{Lacy} {et~al.}(2013){Lacy}, {Ridgway}, {Gates}, {Nielsen}, {Petric},
  {Sajina}, {Urrutia}, {Cox Drews}, {Harrison}, {Seymour}, \&
  {Storrie-Lombardi}}]{lacy13}
{Lacy}, M., {Ridgway}, S.~E., {Gates}, E.~L., {et~al.} 2013, \apjs, 208, 24

\bibitem[{{Lawrence} \& {Elvis}(2010)}]{lawr10}
{Lawrence}, A., \& {Elvis}, M. 2010, \apj, 714, 561

\bibitem[{{Lintott} {et~al.}(2009){Lintott}, {Schawinski}, {Keel}, {van Arkel},
  {Bennert}, {Edmondson}, {Thomas}, {Smith}, {Herbert}, {Jarvis}, {Virani},
  {Andreescu}, {Bamford}, {Land}, {Murray}, {Nichol}, {Raddick}, {Slosar},
  {Szalay}, \& {Vandenberg}}]{lint09}
{Lintott}, C.~J., {Schawinski}, K., {Keel}, W., {et~al.} 2009, \mnras, 399, 129

\bibitem[{{Liu} {et~al.}(2013){Liu}, {Zakamska}, {Greene}, {Nesvadba}, \&
  {Liu}}]{liu13b}
{Liu}, G., {Zakamska}, N.~L., {Greene}, J.~E., {Nesvadba}, N.~P.~H., \& {Liu},
  X. 2013, \mnras, 436, 2576

\bibitem[{{Liu} {et~al.}(2010{\natexlab{a}}){Liu}, {Greene}, {Shen}, \&
  {Strauss}}]{liu10b}
{Liu}, X., {Greene}, J.~E., {Shen}, Y., \& {Strauss}, M.~A. 2010{\natexlab{a}},
  \apjl, 715, L30

\bibitem[{{Liu} {et~al.}(2010{\natexlab{b}}){Liu}, {Shen}, {Strauss}, \&
  {Greene}}]{liu10a}
{Liu}, X., {Shen}, Y., {Strauss}, M.~A., \& {Greene}, J.~E. 2010{\natexlab{b}},
  \apj, 708, 427

\bibitem[{{Liu} {et~al.}(2009){Liu}, {Zakamska}, {Greene}, {Strauss}, {Krolik},
  \& {Heckman}}]{liu09}
{Liu}, X., {Zakamska}, N.~L., {Greene}, J.~E., {et~al.} 2009, \apj, 702, 1098

\bibitem[{{Lupton} {et~al.}(2004){Lupton}, {Blanton}, {Fekete}, {Hogg},
  {O'Mullane}, {Szalay}, \& {Wherry}}]{lupt04}
{Lupton}, R., {Blanton}, M.~R., {Fekete}, G., {et~al.} 2004, \pasp, 116, 133

\bibitem[{{Martin}(2005)}]{mart05}
{Martin}, C.~L. 2005, \apj, 621, 227

\bibitem[{{Mart{\'{\i}}nez-Sansigre} {et~al.}(2006){Mart{\'{\i}}nez-Sansigre},
  {Rawlings}, {Lacy}, {Fadda}, {Jarvis}, {Marleau}, {Simpson}, \&
  {Willott}}]{mart06}
{Mart{\'{\i}}nez-Sansigre}, A., {Rawlings}, S., {Lacy}, M., {et~al.} 2006,
  \mnras, 370, 1479

\bibitem[{{Mateos} {et~al.}(2013){Mateos}, {Alonso-Herrero}, {Carrera},
  {Blain}, {Severgnini}, {Caccianiga}, \& {Ruiz}}]{mate13}
{Mateos}, S., {Alonso-Herrero}, A., {Carrera}, F.~J., {et~al.} 2013, \mnras,
  434, 941

\bibitem[{{Mateos} {et~al.}(2012){Mateos}, {Alonso-Herrero}, {Carrera},
  {Blain}, {Watson}, {Barcons}, {Braito}, {Severgnini}, {Donley}, \&
  {Stern}}]{2012Mateos}
---. 2012, \mnras, 426, 3271

\bibitem[{{Matsuoka} {et~al.}(2014){Matsuoka}, {Strauss}, {Price}, \&
  {DiDonato}}]{matsuoka14}
{Matsuoka}, Y., {Strauss}, M.~A., {Price}, III, T.~N., \& {DiDonato}, M.~S.
  2014, \apj, 780, 162

\bibitem[{{Matsuoka} {et~al.}(2015){Matsuoka}, {Strauss}, {Shen}, {Brandt},
  {Greene}, {Ho}, {Schneider}, {Sun}, \& {Trump}}]{matsuoka15}
{Matsuoka}, Y., {Strauss}, M.~A., {Shen}, Y., {et~al.} 2015, \apj, 811, 91

\bibitem[{{Mazzarella} {et~al.}(2012){Mazzarella}, {Iwasawa}, {Vavilkin},
  {Armus}, {Kim}, {Bothun}, {Evans}, {Spoon}, {Haan}, {Howell}, {Lord},
  {Marshall}, {Ishida}, {Xu}, {Petric}, {Sanders}, {Surace}, {Appleton},
  {Chan}, {Frayer}, {Inami}, {Khachikian}, {Madore}, {Privon}, {Sturm}, {U}, \&
  {Veilleux}}]{mazz12}
{Mazzarella}, J.~M., {Iwasawa}, K., {Vavilkin}, T., {et~al.} 2012, \aj, 144,
  125

\bibitem[{{McCarthy}(1993)}]{mcca93}
{McCarthy}, P.~J. 1993, \araa, 31, 639

\bibitem[{{Mullaney} {et~al.}(2013){Mullaney}, {Alexander}, {Fine}, {Goulding},
  {Harrison}, \& {Hickox}}]{mull13}
{Mullaney}, J.~R., {Alexander}, D.~M., {Fine}, S., {et~al.} 2013, \mnras, 433,
  622

\bibitem[{{Nenkova} {et~al.}(2002){Nenkova}, {Ivezi{\'c}}, \&
  {Elitzur}}]{nenk02}
{Nenkova}, M., {Ivezi{\'c}}, {\v Z}., \& {Elitzur}, M. 2002, \apjl, 570, L9

\bibitem[{{Nenkova} {et~al.}(2008){Nenkova}, {Sirocky}, {Nikutta},
  {Ivezi{\'c}}, \& {Elitzur}}]{nenk08}
{Nenkova}, M., {Sirocky}, M.~M., {Nikutta}, R., {Ivezi{\'c}}, {\v Z}., \&
  {Elitzur}, M. 2008, \apj, 685, 160

\bibitem[{{Nesvadba} {et~al.}(2008){Nesvadba}, {Lehnert}, {De Breuck},
  {Gilbert}, \& {van Breugel}}]{nesv08}
{Nesvadba}, N.~P.~H., {Lehnert}, M.~D., {De Breuck}, C., {Gilbert}, A.~M., \&
  {van Breugel}, W. 2008, \aap, 491, 407

\bibitem[{{Nesvadba} {et~al.}(2006){Nesvadba}, {Lehnert}, {Eisenhauer},
  {Gilbert}, {Tecza}, \& {Abuter}}]{nesv06}
{Nesvadba}, N.~P.~H., {Lehnert}, M.~D., {Eisenhauer}, F., {et~al.} 2006, \apj,
  650, 693

\bibitem[{{Norman} {et~al.}(2002){Norman}, {Hasinger}, {Giacconi}, {Gilli},
  {Kewley}, {Nonino}, {Rosati}, {Szokoly}, {Tozzi}, {Wang}, {Zheng}, {Zirm},
  {Bergeron}, {Gilmozzi}, {Grogin}, {Koekemoer}, \& {Schreier}}]{norm02}
{Norman}, C., {Hasinger}, G., {Giacconi}, R., {et~al.} 2002, \apj, 571, 218

\bibitem[{{Obied} {et~al.}(2016){Obied}, {Zakamska}, {Wylezalek}, \&
  {Liu}}]{obie16}
{Obied}, G., {Zakamska}, N.~L., {Wylezalek}, D., \& {Liu}, G. 2016, \mnras,
  456, 2861

\bibitem[{{P{\^a}ris} {et~al.}(2015){P{\^a}ris}, {Petitjean}, {Aubourg},
  {Ross}, {Myers}, {Streblyanska}, {Bailey}, {Strauss}, {Armengaud},
  {Palanque-Delabrouille}, {Y\`eche}, \& {Hamann}}]{paris15}
{P{\^a}ris}, I., {Petitjean}, P., {Aubourg}, {\'E}., {et~al.} 2015, submitted

\bibitem[{{Pier} \& {Krolik}(1992)}]{pier92}
{Pier}, E.~A., \& {Krolik}, J.~H. 1992, \apj, 401, 99

\bibitem[{{Reid} {et~al.}(2016){Reid}, {Ho}, {Padmanabhan}, {Percival},
  {Tinker}, {Tojeiro}, {White}, {Eisenstein}, {Maraston}, {Ross},
  {S{\'a}nchez}, {Schlegel}, {Sheldon}, {Strauss}, {Thomas}, {Wake}, {Beutler},
  {Bizyaev}, {Bolton}, {Brownstein}, {Chuang}, {Dawson}, {Harding}, {Kitaura},
  {Leauthaud}, {Masters}, {McBride}, {More}, {Olmstead}, {Oravetz}, {Nuza},
  {Pan}, {Parejko}, {Pforr}, {Prada}, {Rodr{\'{\i}}guez-Torres},
  {Salazar-Albornoz}, {Samushia}, {Schneider}, {Sc{\'o}ccola}, {Simmons}, \&
  {Vargas-Magana}}]{reid16}
{Reid}, B., {Ho}, S., {Padmanabhan}, N., {et~al.} 2016, \mnras, 455, 1553

\bibitem[{{Reyes} {et~al.}(2008){Reyes}, {Zakamska}, {Strauss}, {Green},
  {Krolik}, {Shen}, {Richards}, {Anderson}, \& {Schneider}}]{reye08}
{Reyes}, R., {Zakamska}, N.~L., {Strauss}, M.~A., {et~al.} 2008, \aj, 136, 2373

\bibitem[{{Richards} {et~al.}(2003){Richards}, {Hall}, {Vanden Berk},
  {Strauss}, {Schneider}, {Weinstein}, {Reichard}, {York}, {Knapp}, {Fan},
  {Ivezi{\'c}}, {Brinkmann}, {Budav{\'a}ri}, {Csabai}, \& {Nichol}}]{rich03}
{Richards}, G.~T., {Hall}, P.~B., {Vanden Berk}, D.~E., {et~al.} 2003, \aj,
  126, 1131

\bibitem[{{Richards} {et~al.}(2006{\natexlab{a}}){Richards}, {Lacy},
  {Storrie-Lombardi}, {Hall}, {Gallagher}, {Hines}, {Fan}, {Papovich}, {Vanden
  Berk}, {Trammell}, {Schneider}, {Vestergaard}, {York}, {Jester}, {Anderson},
  {Budav{\'a}ri}, \& {Szalay}}]{rich06}
{Richards}, G.~T., {Lacy}, M., {Storrie-Lombardi}, L.~J., {et~al.}
  2006{\natexlab{a}}, \apjs, 166, 470

\bibitem[{{Richards} {et~al.}(2006{\natexlab{b}}){Richards}, {Strauss}, {Fan},
  {Hall}, {Jester}, {Schneider}, {Vanden Berk}, {Stoughton}, {Anderson},
  {Brunner}, {Gray}, {Gunn}, {Ivezi{\'c}}, {Kirkland}, {Knapp}, {Loveday},
  {Meiksin}, {Pope}, {Szalay}, {Thakar}, {Yanny}, {York}, {Barentine},
  {Brewington}, {Brinkmann}, {Fukugita}, {Harvanek}, {Kent}, {Kleinman},
  {Krzesi{\'n}ski}, {Long}, {Lupton}, {Nash}, {Neilsen}, {Nitta}, {Schlegel},
  \& {Snedden}}]{rich06b}
{Richards}, G.~T., {Strauss}, M.~A., {Fan}, X., {et~al.} 2006{\natexlab{b}},
  \aj, 131, 2766

\bibitem[{{Rose} {et~al.}(2015{\natexlab{a}}){Rose}, {Elvis}, {Crenshaw}, \&
  {Glidden}}]{rose15b}
{Rose}, M., {Elvis}, M., {Crenshaw}, M., \& {Glidden}, A. 2015{\natexlab{a}},
  \mnras, 451, L11

\bibitem[{{Rose} {et~al.}(2015{\natexlab{b}}){Rose}, {Elvis}, \&
  {Tadhunter}}]{rose15a}
{Rose}, M., {Elvis}, M., \& {Tadhunter}, C.~N. 2015{\natexlab{b}}, \mnras, 448,
  2900

\bibitem[{{Ross} {et~al.}(2012){Ross}, {Myers}, {Sheldon}, {Y{\`e}che},
  {Strauss}, {Bovy}, {Kirkpatrick}, {Richards}, {Aubourg}, {Blanton}, {Brandt},
  {Carithers}, {Croft}, {da Silva}, {Dawson}, {Eisenstein}, {Hennawi}, {Ho},
  {Hogg}, {Lee}, {Lundgren}, {McMahon}, {Miralda-Escud{\'e}},
  {Palanque-Delabrouille}, {P{\^a}ris}, {Petitjean}, {Pieri}, {Rich}, {Roe},
  {Schiminovich}, {Schlegel}, {Schneider}, {Slosar}, {Suzuki}, {Tinker},
  {Weinberg}, {Weyant}, {White}, \& {Wood-Vasey}}]{ross12}
{Ross}, N.~P., {Myers}, A.~D., {Sheldon}, E.~S., {et~al.} 2012, \apjs, 199, 3

\bibitem[{{Ross} {et~al.}(2015){Ross}, {Hamann}, {Zakamska}, {Richards},
  {Villforth}, {Strauss}, {Greene}, {Alexandroff}, {Brandt}, {Liu}, {Myers},
  {P{\^a}ris}, \& {Schneider}}]{ross15}
{Ross}, N.~P., {Hamann}, F., {Zakamska}, N.~L., {et~al.} 2015, \mnras, 453,
  3932

\bibitem[{{Sanders} {et~al.}(1988){Sanders}, {Soifer}, {Elias}, {Madore},
  {Matthews}, {Neugebauer}, \& {Scoville}}]{sand88}
{Sanders}, D.~B., {Soifer}, B.~T., {Elias}, J.~H., {et~al.} 1988, \apj, 325, 74

\bibitem[{{Schlegel} {et~al.}(1998){Schlegel}, {Finkbeiner}, \&
  {Davis}}]{schl98}
{Schlegel}, D.~J., {Finkbeiner}, D.~P., \& {Davis}, M. 1998, \apj, 500, 525

\bibitem[{{Schneider} {et~al.}(2010){Schneider}, {Richards}, {Hall}, {Strauss},
  {Anderson}, {Boroson}, {Ross}, {Shen}, {Brandt}, {Fan}, {Inada}, {Jester},
  {Knapp}, {Krawczyk}, {Thakar}, {Vanden Berk}, {Voges}, {Yanny}, {York},
  {Bahcall}, {Bizyaev}, {Blanton}, {Brewington}, {Brinkmann}, {Eisenstein},
  {Frieman}, {Fukugita}, {Gray}, {Gunn}, {Hibon}, {Ivezi{\'c}}, {Kent}, {Kron},
  {Lee}, {Lupton}, {Malanushenko}, {Malanushenko}, {Oravetz}, {Pan}, {Pier},
  {Price}, {Saxe}, {Schlegel}, {Simmons}, {Snedden}, {SubbaRao}, {Szalay}, \&
  {Weinberg}}]{schn10}
{Schneider}, D.~P., {Richards}, G.~T., {Hall}, P.~B., {et~al.} 2010, \aj, 139,
  2360

\bibitem[{{Shen} {et~al.}(2011){Shen}, {Liu}, {Greene}, \& {Strauss}}]{shen11b}
{Shen}, Y., {Liu}, X., {Greene}, J.~E., \& {Strauss}, M.~A. 2011, \apj, 735, 48

\bibitem[{{Silk} \& {Rees}(1998)}]{silk98}
{Silk}, J., \& {Rees}, M.~J. 1998, \aap, 331, L1

\bibitem[{{Smee} {et~al.}(2013){Smee}, {Gunn}, {Uomoto}, {Roe}, {Schlegel},
  {Rockosi}, {Carr}, {Leger}, {Dawson}, {Olmstead}, {Brinkmann}, {Owen},
  {Barkhouser}, {Honscheid}, {Harding}, {Long}, {Lupton}, {Loomis}, {Anderson},
  {Annis}, {Bernardi}, {Bhardwaj}, {Bizyaev}, {Bolton}, {Brewington}, {Briggs},
  {Burles}, {Burns}, {Castander}, {Connolly}, {Davenport}, {Ebelke}, {Epps},
  {Feldman}, {Friedman}, {Frieman}, {Heckman}, {Hull}, {Knapp}, {Lawrence},
  {Loveday}, {Mannery}, {Malanushenko}, {Malanushenko}, {Merrelli}, {Muna},
  {Newman}, {Nichol}, {Oravetz}, {Pan}, {Pope}, {Ricketts}, {Shelden},
  {Sandford}, {Siegmund}, {Simmons}, {Smith}, {Snedden}, {Schneider},
  {SubbaRao}, {Tremonti}, {Waddell}, \& {York}}]{smee13}
{Smee}, S.~A., {Gunn}, J.~E., {Uomoto}, A., {et~al.} 2013, \aj, 146, 32

\bibitem[{{Springel} {et~al.}(2005){Springel}, {Di Matteo}, \&
  {Hernquist}}]{spri05}
{Springel}, V., {Di Matteo}, T., \& {Hernquist}, L. 2005, \mnras, 361, 776

\bibitem[{{Stern} {et~al.}(2005){Stern}, {Eisenhardt}, {Gorjian}, {Kochanek},
  {Caldwell}, {Eisenstein}, {Brodwin}, {Brown}, {Cool}, {Dey}, {Green},
  {Jannuzi}, {Murray}, {Pahre}, \& {Willner}}]{ster05}
{Stern}, D., {Eisenhardt}, P., {Gorjian}, V., {et~al.} 2005, \apj, 631, 163

\bibitem[{{Stern} {et~al.}(2012){Stern}, {Assef}, {Benford}, {Blain}, {Cutri},
  {Dey}, {Eisenhardt}, {Griffith}, {Jarrett}, {Lake}, {Masci}, {Petty},
  {Stanford}, {Tsai}, {Wright}, {Yan}, {Harrison}, \& {Madsen}}]{ster12b}
{Stern}, D., {Assef}, R.~J., {Benford}, D.~J., {et~al.} 2012, \apj, 753, 30

\bibitem[{{Strauss} {et~al.}(2002){Strauss}, {Weinberg}, {Lupton}, {Narayanan},
  {Annis}, {Bernardi}, {Blanton}, {Burles}, {Connolly}, {Dalcanton}, {Doi},
  {Eisenstein}, {Frieman}, {Fukugita}, {Gunn}, {Ivezi{\'c}}, {Kent}, {Kim},
  {Knapp}, {Kron}, {Munn}, {Newberg}, {Nichol}, {Okamura}, {Quinn}, {Richmond},
  {Schlegel}, {Shimasaku}, {SubbaRao}, {Szalay}, {Vanden Berk}, {Vogeley},
  {Yanny}, {Yasuda}, {York}, \& {Zehavi}}]{strauss02}
{Strauss}, M.~A., {Weinberg}, D.~H., {Lupton}, R.~H., {et~al.} 2002, \aj, 124,
  1810

\bibitem[{{Tabor} \& {Binney}(1993)}]{tabo93}
{Tabor}, G., \& {Binney}, J. 1993, \mnras, 263, 323

\bibitem[{{Ueda} {et~al.}(2003){Ueda}, {Akiyama}, {Ohta}, \& {Miyaji}}]{ueda03}
{Ueda}, Y., {Akiyama}, M., {Ohta}, K., \& {Miyaji}, T. 2003, \apj, 598, 886

\bibitem[{{Vanden Berk} {et~al.}(2001){Vanden Berk}, {Richards}, {Bauer},
  {Strauss}, {Schneider}, {Heckman}, {York}, {Hall}, {Fan}, {Knapp},
  {Anderson}, {Annis}, {Bahcall}, {Bernardi}, {Briggs}, {Brinkmann}, {Brunner},
  {Burles}, {Carey}, {Castander}, {Connolly}, {Crocker}, {Csabai}, {Doi},
  {Finkbeiner}, {Friedman}, {Frieman}, {Fukugita}, {Gunn}, {Hennessy},
  {Ivezi{\'c}}, {Kent}, {Kunszt}, {Lamb}, {Leger}, {Long}, {Loveday}, {Lupton},
  {Meiksin}, {Merelli}, {Munn}, {Newberg}, {Newcomb}, {Nichol}, {Owen}, {Pier},
  {Pope}, {Rockosi}, {Schlegel}, {Siegmund}, {Smee}, {Snir}, {Stoughton},
  {Stubbs}, {SubbaRao}, {Szalay}, {Szokoly}, {Tremonti}, {Uomoto}, {Waddell},
  {Yanny}, \& {Zheng}}]{vand01}
{Vanden Berk}, D.~E., {Richards}, G.~T., {Bauer}, A., {et~al.} 2001, \aj, 122,
  549

\bibitem[{{Veilleux} \& {Osterbrock}(1987)}]{veil87}
{Veilleux}, S., \& {Osterbrock}, D.~E. 1987, \apjs, 63, 295

\bibitem[{{Villar-Mart{\'{\i}}n} {et~al.}(2011){Villar-Mart{\'{\i}}n},
  {Tadhunter}, {Humphrey}, {Encina}, {Delgado}, {Torres}, \&
  {Mart{\'{\i}}nez-Sansigre}}]{vill11a}
{Villar-Mart{\'{\i}}n}, M., {Tadhunter}, C., {Humphrey}, A., {et~al.} 2011,
  \mnras, 416, 262

\bibitem[{{Whittle}(1985)}]{whit85a}
{Whittle}, M. 1985, \mnras, 213, 1

\bibitem[{{Williams} {et~al.}(2002){Williams}, {Pogge}, \& {Mathur}}]{will02}
{Williams}, R.~J., {Pogge}, R.~W., \& {Mathur}, S. 2002, \aj, 124, 3042

\bibitem[{{Wilson} \& {Heckman}(1985)}]{wils85}
{Wilson}, A.~S., \& {Heckman}, T.~M. 1985, in Astrophysics of Active Galaxies
  and Quasi-Stellar Objects, ed. J.~S. {Miller}, 39--109

\bibitem[{{Wright} {et~al.}(2010){Wright}, {Eisenhardt}, {Mainzer}, {Ressler},
  {Cutri}, {Jarrett}, {Kirkpatrick}, {Padgett}, {McMillan}, {Skrutskie},
  {Stanford}, {Cohen}, {Walker}, {Mather}, {Leisawitz}, {Gautier}, {McLean},
  {Benford}, {Lonsdale}, {Blain}, {Mendez}, {Irace}, {Duval}, {Liu}, {Royer},
  {Heinrichsen}, {Howard}, {Shannon}, {Kendall}, {Walsh}, {Larsen}, {Cardon},
  {Schick}, {Schwalm}, {Abid}, {Fabinsky}, {Naes}, \& {Tsai}}]{wrig10}
{Wright}, E.~L., {Eisenhardt}, P.~R.~M., {Mainzer}, A.~K., {et~al.} 2010, \aj,
  140, 1868

\bibitem[{{Wylezalek} {et~al.}(2016){Wylezalek}, {Zakamska}, {Liu}, \&
  {Obied}}]{wyle16}
{Wylezalek}, D., {Zakamska}, N.~L., {Liu}, G., \& {Obied}, G. 2016, \mnras,
  457, 745

\bibitem[{{York} {et~al.}(2000)}]{york00}
{York}, D.~G., {et~al.} 2000, \aj, 120, 1579

\bibitem[{{Zakamska} \& {Greene}(2014)}]{zaka14}
{Zakamska}, N.~L., \& {Greene}, J.~E. 2014, \mnras, 442, 784

\bibitem[{{Zakamska} {et~al.}(2003){Zakamska}, {Strauss}, {Krolik}, {Collinge},
  {Hall}, {Hao}, {Heckman}, {Ivezi{\'c}}, {Richards}, {Schlegel}, {Schneider},
  {Strateva}, {Vanden Berk}, {Anderson}, \& {Brinkmann}}]{zaka03}
{Zakamska}, N.~L., {Strauss}, M.~A., {Krolik}, J.~H., {et~al.} 2003, \aj, 126,
  2125

\bibitem[{{Zakamska} {et~al.}(2005){Zakamska}, {Schmidt}, {Smith}, {Strauss},
  {Krolik}, {Hall}, {Richards}, {Schneider}, {Brinkmann}, \&
  {Szokoly}}]{zaka05}
{Zakamska}, N.~L., {Schmidt}, G.~D., {Smith}, P.~S., {et~al.} 2005, \aj, 129,
  1212

\bibitem[{{Zakamska} {et~al.}(2006){Zakamska}, {Strauss}, {Krolik}, {Ridgway},
  {Schmidt}, {Smith}, {Heckman}, {Schneider}, {Hao}, \& {Brinkmann}}]{zaka06}
{Zakamska}, N.~L., {Strauss}, M.~A., {Krolik}, J.~H., {et~al.} 2006, \aj, 132,
  1496

\bibitem[{{Zakamska} {et~al.}(2016{\natexlab{a}}){Zakamska}, {Hamann},
  {P{\^a}ris}, {Brandt}, {Greene}, {Strauss}, {Villforth}, {Wylezalek},
  {Alexandroff}, \& {Ross}}]{zaka16b}
{Zakamska}, N.~L., {Hamann}, F., {P{\^a}ris}, I., {et~al.} 2016{\natexlab{a}},
  \mnras, 459, 3144

\bibitem[{{Zakamska} {et~al.}(2016{\natexlab{b}}){Zakamska}, {Lampayan},
  {Petric}, {Dicken}, {Greene}, {Heckman}, {Hickox}, {Ho}, {Krolik},
  {Nesvadba}, {Strauss}, {Geach}, {Oguri}, \& {Strateva}}]{zaka16a}
{Zakamska}, N.~L., {Lampayan}, K., {Petric}, A., {et~al.} 2016{\natexlab{b}},
  \mnras, 455, 4191

\end{thebibliography}

\onecolumngrid
\newpage
\begin{deluxetable}{l|l}
\tablecaption{Entries in the catalog of BOSS type 2 quasars (data model) \label{tab:cat}}
\tablehead{
\colhead{name of parameter} & \colhead{comments}
}
\startdata
plate, fiber, mjd & Spectroscopic identification \\
z & Adopted redshift based on kinematic fits\\ 
ra, dec & Right ascension and declination in decimal degrees\\
oiiiflux, oiiilum, oiiirew & Fluxes (in units of 10$^{-17}\,\ergs$ cm$^{-2}$), luminosities (in units of 10$^{42}\,\ergs$) and rest \\
 & equivalent widths (in \AA) of {[}O {\scriptsize III}{]}$\lambda$5007\AA\ from the complete kinematic fits (Section \ref{subsec:OIII})\\
oiiiw80 & Velocity width containing 80 per cent of {[}O {\scriptsize III}{]}$\lambda$5007\AA\ power (in km s$^{-1}$) \\
nevflux, nevrew, nevsnr & Fluxes, rest equivalent widths and signal-to-noise ratios of {[}Ne {\scriptsize V}{]}$\lambda$3426\AA\ from \\
 & single-Gaussian fits \\
magu, magg, magr, magi, magz & Model SDSS magnitudes (corrected for
\citealt{schl98} extinction)\\
emagu, emagg, emagr, emagi, emagz & Errors on the model SDSS magnitudes \\
w1, w2, w3, w4 & WISE catalog magnitudes (in Vega system) \\
ew1, ew2, ew3, ew4 & Errors on the WISE magnitudes \\
lum5, lum12 & Rest-frame luminosities $\nu L_{\nu}$ at 5 and 12\micron\ calculated from piece-wise interpolation \\
 & between WISE fluxes (in units of 10$^{42}\,\ergs$) \\
select & A string value indicating how the object was selected: possible values are \\
 & `Lowz' (Section \ref{ssec:lowz}), `Highz' (Section
\ref{ssec:highz}), `Wrongz1', `Wrongz2' and `Zwarning' (Section \ref{ssec:wrongz}) \\
unique & A string value set to `unique' if the object is making its first or only appearance in the catalog; \\
 & if not, the flag is set to the spectroscopic ID of the first appearance of the source \\
 & in the format `pppp-ffff-mmmmm' \\
\enddata
\end{deluxetable}

\begin{deluxetable}{l|l}
\tablecaption{Kinematic parameters of {[}O$\,${\scriptsize III}{]}$\lambda$5007\AA\ (data model) \label{tab:kin}}
\tablehead{
\colhead{name of parameter} & \colhead{comments}
}
\startdata
plate, fiber, mjd & Spectroscopic identification \\
z & Adopted redshift \\ 
amp1, vel1, sig1 & Amplitude (in units of $10^{-17}\,\ergs$ cm$^{-2}$ \AA$^{-1}$), velocity offset (in km s$^{-1}$) and velocity dispersion \\ 
  & (in km s$^{-1}$) of the first Gaussian component as measured in the frame placed at \\
  & the adopted redshift \\
amp2$-$4, vel2$-$4, sig2$-$4 & Same as above for additional Gaussian components; 0 amplitude indicates the component \\
 & is not required by the fit \\
fwhm, fwqm & Full width at half maximum and at quarter maximum of the line profile, in km s$^{-1}$ \\
w50, w80, w90 & Velocity widths containing 50 per cent, 80 per cent and 90 per cent of the line profile power, in km s$^{-1}$ \\
relasym, r9050 & Dimensionless relative asymmetry $R$ and kurtosis parameter $r_{9050}$ \\
dp & Double-peaked candidate flag: 0 -- no, 1 -- visual inspection, 2 -- profile minimum and visual inspection \\ 
\enddata
\tablecomments{These data are provided in two on-line FITS tables: one
  for the new catalog of SDSS-III type 2 quasars presented here and
  one for the 568 objects with \oiii\ kinematics calculated by
  \citet{zaka14}. The files can be found at \url{http://zakamska.johnshopkins.edu/data.htm}. } 
\end{deluxetable}

\end{document}